\documentclass[a4paper,12pt]{article}

\usepackage{hyperref}
\usepackage{epsfig}
\usepackage{amssymb}
\usepackage{setspace}

\usepackage{amsmath}
\usepackage{amssymb}
\usepackage{graphicx}
\usepackage{geometry}
\usepackage{subfigure}
\usepackage{url}

\newcommand{\mrm}[1]{\mbox{\rm #1}}
\newcommand{\be}{\begin{equation}}
\newcommand{\ee}{\end{equation}}
\newcommand{\br}{\begin{eqnarray}}
\newcommand{\bea}{\begin{eqnarray}}
\newcommand{\eea}{\end{eqnarray}}
\newcommand{\er}{\end{eqnarray}}
\newcommand{\ba}{\begin{array}}
\newcommand{\ea}{\end{array}}
\newcommand{\bi}{\begin{itemize}}
\newcommand{\ei}{\end{itemize}}
\newcommand{\bn}{\begin{enumerate}}
\newcommand{\en}{\end{enumerate}}
\newcommand{\bc}{\begin{center}}
\newcommand{\ec}{\end{center}}

\newcommand{\Eq}[1]{Eq.~(\ref{#1})}

\def\l{\lambda}
 
  \def\b{\beta}
 \def\({\left(}
 \def\){\right)}
 \def\[{\left[}
 \def\]{\right]}
 \def\m{\mu}

  \def\cL{\mathcal{L}}
\def\mHu{{m_{h_u}}}
 \def\mHd{{m_{h_d}}}
 \def\mHu2{{m_{h_u}^2}}
 \def\mHd2{{m_{h_d}^2}}
 \def\nn{\nonumber}
 \def\a{\alpha}
 \def\half{\frac{1}{2}}

\newcommand{\hc}[1]{#1^\dagger}
\newcommand{\vev}[1]{\langle #1 \rangle}

\def\gappeq{\mathrel{\rlap {\raise.5ex\hbox{$>$}}
{\lower.5ex\hbox{$\sim$}}}}

\def\lappeq{\mathrel{\rlap{\raise.5ex\hbox{$<$}}
{\lower.5ex\hbox{$\sim$}}}}

\begin{document}
\pagestyle{empty}
\begin{flushright}
CERN-PH-TH-2012-078 \\
\end{flushright}
\vspace*{10mm}
\begin{center}
{\LARGE {\bf  Fermiophobic Higgs boson and supersymmetry}} \\
\vspace*{1.5cm}
{\large
 {\bf E. Gabrielli$^{{a,b}}$}, {\bf K. Kannike$^{{a,c}}$},  {\bf B. Mele$^{{d}}$}, {\bf A.~Racioppi$^{a}$},
  {\bf M.~Raidal$^{a,e,f}$}}
\vspace{0.3cm}

{\it
{ (a) NICPB, Ravala 10, Tallinn 10143, Estonia}  \\[1mm]
{ (b) INFN, Sezione di Trieste, c/o Dipartimento di Fisica, Universit\` a
di Trieste,\\ Via Valerio 2, I - 34127 Trieste, Italy }  \\[1mm]
{ (c)  Scuola Normale Superiore and INFN, Piazza dei Cavalieri 7, 56126 Pisa, Italy}  \\[1mm]
{ (d) INFN, Sezione di Roma, c/o Dipartimento di Fisica, Universit\`a di Roma ÒLa SapienzaÓ, \\Piazzale A. Moro 2, I-00185 Rome, Italy} \\[1mm]
{\it  (e) Institute of Physics, University of Tartu, Estonia}\\[1mm] }
 {\it (f) CERN, Theory Division, CH-1211 Geneva 23, Switzerland}\\[1mm]

\vspace*{2cm}
{\bf ABSTRACT} \\
\end{center}
\vspace*{5mm}
\noindent
If a light Higgs boson with mass 125~GeV is fermiophobic,
or partially fermiophobic, then the MSSM is excluded.
The minimal supersymmetric fermiophobic  Higgs scenario can naturally be formulated in the context of the NMSSM that admits $Z_3$ discrete symmetries.
In the fermiophobic NMSSM, the SUSY naturalness criteria are relaxed by a factor $N_c y_t^4/g^4 \sim 25,$ removing the little hierarchy problem and
allowing sparticle masses to be naturally of order 2--3~TeV.
This scale motivates  wino or higgsino dark matter. The SUSY flavour and CP problems as well as  the constraints on sparticle and Higgs boson masses
from $b\to s\gamma,$ $B_s\to \mu\mu$ and direct LHC searches are relaxed in fermiophobic NMSSM.
The price to pay is that a new, yet unknown, mechanism must be introduced to generate fermion masses.
 We show that in the fermiophobic NMSSM the radiative Higgs boson branchings to $\gamma\gamma$, $\gamma Z$
 can be modified compared to the fermiophobic and ordinary standard model predictions, and fit present collider data better.
Suppression of dark matter scattering off nuclei explains the absence of signal in XENON100.

\vspace*{1.5cm}
\noindent

\begin{flushleft}
March  2012
\end{flushleft}
\vfill\eject

\setcounter{page}{1}
\pagestyle{plain}


\section{Introduction and motivation}

The TeVatron \cite{gammagammaTevatron,bbTeVatron} and LHC experiments
presented their new and updated results~\cite{Agg,Cgg,AZZ,CZZ,CWW,WW,tautau,bbCMS,Acomb,Ccomb}  on  searches for the Higgs boson~\cite{Higgs}
at the Moriond 2012 conference~\cite{moriond}. While on average the data  is consistent with the standard model (SM) Higgs boson with
mass 125~GeV, interesting anomalies start to emerge that may signal unexpected new physics in the Higgs sector. The most interesting of them
is a local $3\sigma$ level excess in searches for the fermiophobic (FP) Higgs boson \cite{fermiophobic:theory} in $\gamma\gamma$ final states both in the ATLAS and
CMS experiments \cite{fermiophobic}. This signals that there is an anomalously large contribution in the observed $\gamma\gamma$
excess  coming from  the  vector-boson fusion (VFB)  Higgs production mechanism.
Indeed, the relative weight of the latter and the associate production with $W,Z$ (VH) is  enhanced
 with respect to the SM dominant gluon-gluon fusion channel (ggF)  in the FP high-$p_{T}$ selections applied by the CMS and ATLAS.
This anomaly is accompanied by a deficit of $WW^*$ compared with the SM in all experiments.

The Higgs boson mass $M_h\approx 125$~GeV is peculiar in several ways. In the context of FP Higgs boson,
there is an accident that at the 7--8~TeV LHC  the Higgs boson signal rate in the $\gamma\gamma$ channel, $\sigma\times BR,$
happens to be equal to the SM one in the vicinity of this Higgs mass value~\cite{Gabrielli:2012yz}. This is the reason why the LHC inclusive $\gamma\gamma$
excess is consistent with the FP Higgs boson. At the same time, the signal rates for other gauge boson channels, $WW^*,$ $ZZ^*,$ $Z\gamma,$
are predicted for be 40-50\% suppressed compared to the SM~\cite{Gabrielli:2012yz}.  The $M_h\approx 125$~GeV Higgs boson is peculiar  also
 because, for  a SM-like Higgs of that mass,   branching fractions for many  decay channels  are measurable at the LHC. Therefore, the LHC is, in principle, able
to determine the nature and properties of the 125~GeV Higgs boson.

Motivated by these results we performed a global fit to all available collider data to determine which Higgs boson scenario
is currently favoured~\cite{Giardino:2012ww}, improving and extending similar pre-Moriond fits~\cite{Carmi:2012yp}.
A purely FP Higgs boson gives a fit to present data almost as good as the SM one,  but
with very different predictions for the signal rates at the LHC. A partially FP Higgs boson, however, gives a significantly better fit
to current data than the SM~\cite{Giardino:2012ww}. This is because the FP Higgs qualitatively describes the observed anomalies in the data
 correctly, although it predicts larger signal rates than observed. Small additional branching fractions into fermionic channels, for example into $b\bar b,$
decrease the overall rate and improve the fit to data significantly.

Partial fermiophobia is exactly what is expected to happen when considering the FP Higgs boson scenario as an effective low energy theory in
the context of quantum field theory~\cite{Gabrielli:2010cw}. Because the SM fermions are massive, at loop level non-vanishing
Yukawa couplings are generated even if at some high new physics scale
the Higgs boson was initially purely fermiophobic. As long as the fermion mass generating mechanism
is unknown, the induced Yukawa couplings represent an uncertainty in the low energy FP Higgs model.
Present data  suggest that the Higgs boson might be partially fermiophobic~\cite{Gabrielli:2012yz,Gabrielli:2010cw,Berger:2012sy}.

The 125~GeV Higgs boson is also unsatisfactory, because this mass is below the SM vacuum stability bound in case new physics appears only
above the  scale of gauge coupling unification~\cite{EliasMiro:2011ex}. The vacuum stability can be made consistent with the
125~GeV Higgs mass~\cite{scalar:dm:vacuum:stab} by extending the scalar sector with dark matter
candidates~\cite{realsinglet,inert}. But the simplest solution is provided by
(partial) fermophobia. Because the vacuum instability is caused by the large top Yukawa coupling in the SM, reducing its value makes the
hinted Higgs mass compatible with the GUT scale.

It is well known that the existence of  the Higgs boson rises a question of why the electroweak scale is so much smaller than the Planck scale.
The most elegant solution to that problem is given by supersymmetry (SUSY). However, direct and indirect collider bounds, cosmological
dark matter abundance and constraints from dark matter direct detection experiments together with a Higgs boson mass of 125~GeV
impose  stringent constraints on SUSY scale in most popular SUSY models -- the
MSSM~\cite{Djouadi:2005gj,MSSM:Higgs} and NMSSM~\cite{nmssm,NMSSM:etc}.
The constraint $M_{\rm SUSY}>\mrm{1~TeV}\gg M_Z$ reintroduces severe fine tuning to theory,
known as the little hierarchy problem~\cite{Barbieri:2000gf}, that
 makes SUSY as a solution to the hierarchy problem unnatural.

If the Higgs boson turns out to be fermiophobic, some SUSY models are in even more serious trouble. We show that  fermiophobia
and a Higgs boson mass at 125~GeV together exclude all versions of the MSSM  independently of any model detail.
This is because in the MSSM the upper bound on the tree level Higgs boson mass is $M_Z$, and large radiative corrections,
dominated by stop contributions, are needed to reach 125~GeV. Fermiophobia removes the dominant  stop loops as they are
induced by Yukawa couplings. There might be large trilinear scalar couplings, the $A$-terms, but their contribution to $M^2_h$ is
negative. Dimensionful trilinear couplings may trigger electroweak symmetry breaking~\cite{radiative:EWSB}
via a  dimensionful  Coleman-Weinberg~\cite{Coleman} mechanism,
but cannot increase the Higgs boson mass.

Nevertheless, SUSY models with additional tree level contributions to the Higgs boson mass, such as the NMSSM,
are viable fermiophobic SUSY theory candidates. In fact fermiophobia can cure some SUSY problems and make it more
compatible with experimental data. Firstly, the fine tuning problem of SUSY, also coming from loop contributions to the Higgs
boson mass squared, is now induced by gauge couplings, improving the fine tuning by a factor of  $N_c y_t^4/g^4 \sim 25.$
This improvement completely removes the little hierarchy problem. SUSY masses of order 2-3~TeV become completely
natural. Allowing for some fine tuning, even the split SUSY~\cite{ArkaniHamed:2004fb}  with a very heavy scalar sector becomes viable.
This would explain why no sparticles have been discovered by the LHC up to now~\cite{cmssusy,atlassusy}.
Secondly, SUSY flavour and CP problems are improved
by removing (or decreasing) the Yukawa couplings and by
allowing also squark and slepton masses to be at a few-TeV scale.
Thirdly, the additional Higgs bosons can be light.
For example, the absence of constraints from $b\to s \gamma$ and $B_s\to \mu\mu$~\cite{bsmumu} allows the charged Higgs
boson to be light, opening again the possibility for its discovery at the LHC. Fourthly, the SUSY fermion sector
may be either light or heavy. While gluinos could have been abundantly produced at the LHC if they were very light,
for colourless fermions there would exist only collider bounds from LEP and Tevatron.
Fifthly, the constraints on dark matter are relaxed. Because neutralino elastic scattering off nuclei is dominated
by tree level Higgs boson exchange, this process would be suppressed and the prospects for dark matter discovery at the XENON100 are
decreased in this scenario~\cite{Hisano:2010fy}.  As the SUSY scale could now be  large,
higgsino or wino relic abundance would become a natural explanation to the dark matter of the Universe.

The obvious question in any FP Higgs boson scenario is what is the alternative mechanism for generating the observed
fermion masses. Because the top quark mass is so large, it cannot be generated radiatively.  The most plausible
scenario for generating such large fermion masses is strong dynamics above the electroweak scale~\cite{Heff}.
In such a scenario both the composite Higgs boson fermiophobia and fermion masses might  originate from the same new physics.
A generic prediction of strong electroweak symmetry breaking scenarios, including composite Higgs models, is
the appearance of new resonances at 2--3~TeV. In the following, we assume that such or any other new physics scenario
above the electroweak scale generates the top quark mass.

In this work we formulate a FP NMSSM as a minimal FP SUSY model. Originally the NMSSM was constructed to
solve the $\mu$ problem of the MSSM and to have an additional contribution to the masses of the Higgs bosons.
To achieve that, an additional $Z_3$ symmetry is usually imposed on the NMSSM. We show that choosing quantum numbers of
this symmetry appropriately, superpotential Yukawa terms may be forbidden in the NMSSM.  In this case, all the
pros and contras of FP SUSY discussed above apply also to the FP NMSSM. In our phenomenological study of the model
we concentrate on radiatively induced Higgs boson decays $h\to \gamma\gamma$ and $h\to Z\gamma.$
This choice is motivated by the fact that SUSY effects most easily show up in loop level processes. Our aim is to study
whether the FP SUSY  Higgs boson gives a better or worse phenomenological fit to the LHC data than the FP SM Higgs boson.
We find that the new SUSY contribution can enhance or reduce the $\gamma\gamma$ and $Z\gamma$ rates as much
as 50\% compared to the FP SM Higgs depending on the sign of the $\mu$ parameter. Because the data prefers smaller
rates~\cite{Giardino:2012ww},  the FP SUSY Higgs can give a better fit to data than the SM.

We note that in the MSSM  the $WW^*$ and $ZZ^*$ rates can be reduced at tree level compared to the SM.
At the same time, the decay $h\to \gamma\gamma$ is  dominated by the $W$-boson loop, introducing a correlation between the two processes.
Should the observed deficit in $WW^*$ persist together with the $\gamma\gamma$ excess, new physics beyond the MSSM would be required.
A FP NMSSM might be a good candidate for such new physics.

The paper is organized as follows. In section 2 we study fermiophobic Higgs scenarios in SUSY. In section 3 we study the radiative decays
$h\to \gamma\gamma$ and $h\to Z\gamma$ in a specific parameter region of the FP NMSSM model.
In section 4 we add some discussion and conclude in section 5.

\section{Fermiophobic supersymmetry}

To formulate a supersymmetric FP Higgs boson theory,  the first attempt should be made in the MSSM.
However, we are going to show that the Higgs boson mass $M_h\approx 125$~GeV is by far too large to be generated
in the FP MSSM since  loop corrections from the top Yukawa coupling are absent. The dominant SUSY loop contribution to the Higgs mass
 in the FP MSSM comes from the large trilinear $A$-term, but this contribution is always negative. Thus the FP MSSM is definitely
excluded on phenomenological grounds.
In order to rescue supersymmetry, we show that the NMSSM offers a natural framework to formulate a supersymmetric FP Higgs scenario
consistent with experimental results.

\subsection{Fermiophobic MSSM}
The well known MSSM superpotential is
\begin{equation}
  \mathcal{W} = y_u Q H_u u^c + y_d H_d Q d^c + y_e H_d L e^c + \mu H_u H_d,
  \label{WMSSM}
\end{equation}
where $y_u,$ $y_d$ and $y_e$ are the up quark, down quark and charged lepton Yukawa couplings.
The LHC hints for a FP Higgs imply that at least the third generation Yukawa couplings must be
strongly suppressed compared to their SM values so that the production mechanism $gg\to h$ and the decay channels $h\to b\bar b,$  $h\to \tau\bar \tau$
become subdominant compared  to the gauge boson processes. Following this indication, in this paper we
study its implications without trying to explain the suppression of Yukawa couplings in the context of MSSM.
Thus, for simplicity,  we just take $y_u=y_d=y_e=0.$

In the MSSM the tree level Higgs boson mass has the well known upper bound $M_h^{\rm tree}<M_Z.$
This comes from the fact that in SUSY the Higgs quartic couplings are generated by gauge couplings via the $D$-terms.
As the Higgs boson quartic coupling is the only free parameter in the SM Higgs sector, in the MSSM there is no
freedom to tune the tree level Higgs boson mass. To be consistent with experimental data, in the MSSM
very large positive loop corrections to $M_h^2$ must be generated.  Those loop corrections are dominated by
top squark contributions that are induced by the top Yukawa coupling $y_t$~\cite{Carena:1995bx},
\be
 \Delta M_h^2 = 3 y_t^4 \frac{v^2 \sin^4\b}{8 \pi^2} \[ \log\frac{M_S^2}{m_t^2} + \frac{X_t^2}{M_S^2} \( 1- \frac{X_t^2}{12 M_S^2}\)  \],
 \label{eq:eps}
\ee

where $M_S$ is the average stop mass, $\tan\beta=v_u/v_d,$  $v^2=v_u^2+v_d^2,$  and $X_t$ is the stop mass mixing parameter
\be
 X_t = A_t - \m \cot\b = \frac{a_t}{y_t} - \m \cot\b,
 \label{eq:Xt}
\ee
where $a_t$ is the trilinear coupling of the soft term $a_t \tilde Q H_u\tilde u^c$.
In order to achieve the Higgs boson mass
indicated by the LHC experiments, $M_h\approx 125$~GeV, \Eq{eq:eps} implies that the stop masses must exceed TeV scale.

In the FP MSSM the dominant stop contribution is absent since we take $y_t \to 0.$  However,  loops induced by very large trilinear soft interaction
$a_t \tilde Q H_u {\tilde u}^c$ with $a_t>1$~TeV are still allowed in general MSSM even in the absence of Yukawa couplings.
In the FP limit we obtain
\be
\Delta M_h^2 = - \frac{3 v^2 \sin^4\b}{8 \pi^2} \frac{a_t^4}{12 M_S^4},
\ee
that  is negative.  The dimensionful couplings like the $A$-terms
 may be used to generate negative Higgs mass terms radiatively, thus generating dynamical breaking of electroweak symmetry~\cite{radiative:EWSB},
 but they do not increase the Higgs mass prediction in the MSSM. This is because the Higgs boson quartic self coupling is fixed by
 the $D$-term that gives the upper bound. As the chargino loop contributions to the Higgs boson mass are of order few GeV,
 in the FP MSSM the Higgs boson mass 125~GeV is not achievable. Independently of model details, the FP MSSM is excluded by the Higgs
 boson mass.

\subsection{Fermiophobic NMSSM}

NMSSM is the next to minimal supersymmetric standard model whose particle content is extended by a gauge singlet chiral superfield $S$ (for reviews and references therein see \cite{nmssm}).
The original motivation for the NMSSM was to explain why the MSSM superpotential parameter $\mu H_u H_d$ is of the same order as the
soft SUSY breaking parameters. In addition, in the NMSSM the Higgs bosons obtain tree level mass not determined by the $D$-terms, thus allowing
larger Higgs masses than $M_Z.$ To achieve those goals, usually the most general NMSSM is constrained by imposing an additional $Z_3$
symmetry in addition to the $R$-parity. Those properties make the NMSSM our prime candidate for the minimal FP SUSY model.
The superpotential of the FP NMSSM that we would like to obtain is given by
\begin{equation}
  \mathcal{W} = \lambda S H_u H_d + \frac{\kappa}{3} S^3,
  \label{W}
\end{equation}
together with the following soft SUSY breaking terms
\be
 \cL_{\text{soft}} = - \( \mHu2 h_u^\dagger h_u + \mHd2 h_d^\dagger h_d + m_s^2 s^\dagger s \) - \( a_\l s h_u h_d + \frac{1}{3} a_k s^3 + {\rm h.c.} \),
 \label{soft}
\ee
where $s$ stands for the scalar component of the singlet chiral superfield $S$.  Thus we have to forbid the
$\mu H_u H_d$, $S$, $S^2$, $y_u Q H_u u^c$, $y_d H_d Q d^c$, $y_e H_d L e^c$ terms by imposing an additional $Z_N$ symmetry and appropriately
choosing the $Z_N$ charges to satisfy the following constraints
\begin{align}
X_Q + X_{H_u} + X_{u^c} \neq 0 & \mod N, \\
X_Q + X_{H_d} + X_{d^c} \neq 0 & \mod N, \\
X_L + X_{H_d} + X_{e^c} \neq 0 & \mod N, \label{eq:forbid:lepton:yukawa} \\
X_{H_u} + X_{H_d} \neq 0 & \mod N, \\
X_S \neq 0 & \mod N, \\
2 X_S \neq 0 & \mod N, \\
3 X_S = 0 & \mod N, \\
X_S + X_{H_u} + X_{H_d} = 0 & \mod N.
\end{align}
One could choose $X_L = X_Q$ and $X_{e^c} = X_{d^c}$, then \eqref{eq:forbid:lepton:yukawa} would be superfluous.

For $Z_3$, if $X_{H_u} = X_{H_d} = X_S = 1$ and $X_{\rm fermion} = 0$, the above equations are satisfied.
In addition, the lowest order Yukawa couplings can be generated via  $d=6$ operator $\frac{\vev{\hc{H_d}} \vev{H_d}}{\Lambda^2} Q H_d d^c$,
where $\Lambda$ is the scale of new physics. This demonstrates that small but non-vanishing Yukawa couplings should exist also in the FP Higgs scenarios.
On the other hand,  if $X_{H_u} = X_{H_d} = X_S = 1$ and $X_{\rm fermion} = 2$, the above equations are also satisfied, but the lowest order Yukawas would be generated as e.g. $\frac{\vev{S}}{\Lambda} Q H_d d^c$.
These charges satisfy two additional equations
\begin{align}
  X_Q + X_{H_u} + X_{u^c} + X_S &= 0, \\
  X_Q + X_{H_d} + X_{d^c} + X_S &= 0.
\end{align}

This could possibly generate the Yukawa couplings for the first two generations but not for the top quark. Therefore we have to assume that
the significant amount of third generation fermion masses should come from some additional mechanism. The prime candidate for such a
mechanism is some strong dynamics above 2--3~TeV scale.

The $Z_{3}$ symmetry could come from the breaking of an $U(1)'$ \cite{u1tozn}. In this case one has to keep in mind the possibility of discrete gauge anomalies \cite{discreteanomalies}. In the NMSSM these have been considered in \cite{discr:gauge:nmssm}. If one chooses $X_{\rm fermion} = 0$, the anomaly constraints can be simply evaded.

Breaking of the $Z_3$ symmetry in the early universe could create a problem with the domain walls that produce an anisotropy in the CMB and ruin nucleosynthesis \cite{walls}. The problem can be solved by allowing for radiative generation of small renormalisable $Z_3$ breaking terms \cite{domain:wall:solution}.

\section{Phenomenology of FP NMSSM Higgs bosons at the LHC}

Our aim in this section is to study radiatively induced decays   $h\to \gamma\gamma$ and $h\to Z\gamma$
of the SM-like FP Higgs boson in the context of NMSSM. If there are light superparticles or light additional Higgs bosons,
their effects are first expected to show up in loop level processes. However, before proceeding with this study, we have to
show that the FP NMSSM is a viable model. Therefore we start studying the FP NMSSM Higgs sector.
We do not attempt to scan the full parameter space of the model.
Instead, we start by fixing the
parameters in the Higgs sector to one particularly interesting
point with $\tan\beta=1$ and decoupled CP-even singlet, that
allows for two following Higgs scenarios: $(i)$ the 125~GeV excess is due to
the lightest CP-even Higgs boson, the remaining  neutral Higgs
bosons are too heavy and have no direct decays to two gauge
bosons; $(ii)$  the 125~GeV excess is due to the next-to-lightest
CP-even Higgs boson, the lightest one has no direct decays into
two gauge bosons and remains invisible at the LHC.
\\
Here, we will focus on scenario $(i)$.
Then, we relax
the condition $\tan\beta =1$ and analyze the impact of a
$\tan\beta\neq 1$ value, in the approximation in which the scalar
singlet is very heavy.
We finally show how the FP NMSSM can be distinguished from the FP SM by studying the radiatively induced Higgs boson decays.

\subsection{Scalar potential and masses}

 The FP NMSSM scalar potential derived from \Eq{W}, \Eq{soft} and also from  the $D$-term contributions is
\bea
 V &=& \left(m_{h_u}^2 + \left|\lambda s\right|^2\right) \left(\left|h_u^0\right|^2 + \left|h_u^+\right|^2\right)
      +\left(m_{h_d}^2 + \left|\lambda s\right|^2\right) \left(\left|h_d^0\right|^2 + \left|h_d^-\right|^2\right) \nn \\
    &&+m_{s}^2 |s|^2 +\big( a_\l \left(h_u^+ h_d^- - h_u^0 h_d^0\right) s  +\frac{1}{3} a_k s^3 + \mathrm{h.c.}\big) \nn \\
    &&+\left|\lambda \left(h_u^+ h_d^- - h_u^0 h_d^0\right) + k s^2 \right|^2 \nn \\
    &&+\frac{g_1^2+g_2^2}{8}\left(\left|h_u^0\right|^2 + \left|h_u^+\right|^2 - \left|h_d^0\right|^2 - \left|h_d^-\right|^2\right)^2
      +\frac{g_2^2}{2}\left|h_u^+ h_d^{0*} + h_u^0 h_d^{-*}\right|^2.
      \label{V}
\eea
We suppose for simplicity that all the parameters in \Eq{V} are real.  We have checked that the potential is always bounded from below.
Supposing that only the real parts of Higgs bosons can get  vacuum expectation values (VEVs) different from
zero and parametrizing the fields as
\be
  h_{d}^0 = \frac{1}{\sqrt2} \( v_d+h_{dR}^0 + i h_{dI}^0 \), \qquad
  h_{u}^0 = \frac{1}{\sqrt2} \( v_u+h_{uR}^0 + i h_{uI}^0 \), \qquad
  s = \frac{1}{\sqrt2} \( v_S+s_R + i s_I \),
\ee
we get the following equations for the stationary points
\bea
   && v \Big(-4 v_S \sin\b \left(\sqrt{2} a_\l+k \l^2 v_S\right)
   +\cos\b \left(-v^2 \sin^2\b
   \left(g_1^2+g_2^2-4 \l^2\right)+8 \mHd2+4 \l^2  v_S^2\right) \nn\\
   &&\qquad +v^2 \cos^3\b  \left(g_1^2+g_2^2\right)\Big) = 0, \label{eq:minHd}\\
   && v \Big(-4 v_S \cos\b \left(\sqrt{2} a_\l+k \l^2 v_S\right)
   +\sin\b \left(v^2 \sin^2\b \left(g_1^2+g_2^2\right)+8 \mHu2+4 \l^2 v_S^2\right) \nn\\
   &&\qquad -v^2 \sin\b \cos^2\b    \left(g_1^2+g_2^2-4 \l^2\right)\Big) = 0, \label{eq:minHu} \\
   && v_S \left(\sqrt{2} a_kv_S+\l^2 \left(2 k^2 v_S^2+v^2\right)+2 m_S^2\right)
   -\frac{1}{4} v^2 \sin(2\beta) \left(\sqrt{2} a_\l+2 k \l^2 v_S\right) = 0. \label{eq:minS}
\eea
First of all we must avoid that $v_u=v_d=v_S=0$ is a minimum of
the potential. This can be done by requiring
\be
 m_{h_u}^2 m_{h_d}^2 m_{s}^2 < 0.
\ee
Now let us give a look at the Hessian matrix in the minimum. In the
basis $(h_{dR}^0,h_{uR}^0,s_R)$ we have for the CP-even Higgs bosons
\be M_S^2 = \left(
\begin{array}{ccc}
 M_{S,11}^2 &  M_{S,12}^2 &  M_{S,13}^2  \\
 \dots      &  M_{S,22}^2 &  M_{S,23}^2  \\
 \dots      & \dots       & M_{S,33}^2
\end{array}
\right),
 \ee
where
\bea
 M_{S,11}^2 &=& \mHd2+\frac{v_S^2 \l^2}{2}+\frac{1}{8}    v^2 \left(g_1^2+g_2^2+2 \l^2+2    \left(g_1^2+g_2^2-\l^2\right) \cos(2\beta)\right), \\
 M_{S,22}^2 &=& \mHu2+\frac{v_S^2 \l^2}{2}+\frac{1}{8} v^2  \left(g_1^2+g_2^2+2 \l^2-2  \left(g_1^2+g_2^2-\l^2\right) \cos(2\beta)\right), \\
 M_{S,33}^2 &=& m_S^2+3 k^2 v_S^2 +\sqrt{2} a_kv_S+v^2 \left(\frac{\l^2}{2}-k \l^2 \cos\b \sin\b\right), \\
 M_{S,12}^2 &=& \frac{1}{8} \left(-g_1^2-g_2^2+4 \l^2\right) \sin(2\beta)    v^2-\frac{1}{2} k v_S^2 \l-\frac{a_\l    v_S}{\sqrt{2}}, \\
 M_{S,13}^2 &=&  v v_S \left(\l^2 \cos\b-k \l \sin\b \right)-\frac{a_\l v \sin\b}{\sqrt{2}},\\
 M_{S,23}^2 &=&  v v_S \left(\l^2 \sin\b-k \l    \cos\b\right)-\frac{a_\l v \cos\b}{\sqrt{2}}.
\eea
So far the results have been general. However,
we can see from the mass matrix  that there is a choice of the parameters that allows no
mixing between $s$ and $h_{u,d}^0,$
\bea
\tan\b &=&1,  \label{eq:nomixingSH1} \\
 k &=& \l,  \label{eq:nomixingSH2} \\
 a_\l &=&  \label{eq:nomixingSH3} 0 .
\eea
Notice that $\tan\beta=1$ is allowed in this model because no constraints occur from the scalar potential minimization nor from the Yukawa sector.
Therefore this choice is the most natural one.
Adopting, for simplicity,  the choice in eqs. (\ref{eq:nomixingSH1})-(\ref{eq:nomixingSH3}) and requiring, of course,
$v_u\neq0,v_d\neq0,v_S\neq0,$ the minimization equations read
\bea
 \left(4 \mHd2+\l^2 v^2\right) &=& 0, \label{eq:minHdnew}\\
 \left(4 \mHu2+\l^2 v^2\right) &=& 0, \label{eq:minHunew} \\
 \frac{a_kv_S^2}{\sqrt{2}}+m_S^2 v_S+\l^2 v_S^3 &=& 0, \label{eq:minSnew}
\eea
leading to the CP-even neutral Higgs boson mass matrix
\be
 M_S^2 = \left(
\begin{array}{ccc}
 \frac{1}{2} \left(M_Z^2+v_S^2 \l^2\right)           & \frac{1}{2} \left((v-v_S) (v+v_S) \l^2-M_Z^2\right) & 0 \\
 \frac{1}{2} \left((v-v_S) (v+v_S) \l^2-M_Z^2\right) & \frac{1}{2} \left(M_Z^2+v_S^2 \l^2\right)           & 0 \\
                                        0            &                            0                        & 2 v_S^2 \l^2+\frac{a_kv_S}{\sqrt{2}}
\end{array}
\right),
\label{eq:MSnomix}
 \ee
where $M_Z^2=\frac{1}{4} \(g_1^2+g_2^2\) v^2$.
The corresponding eigenvalues
\bea
 M_{h^0_1}^2 &=& \frac{\lambda^2 v^2}{2} ,\\
 M_{h^0_2}^2 &=& M_Z^2+\frac{1}{2} \lambda^2 \left( 2 v_S^2 - v^2 \right) ,\\
 M_{s_R}^2 &=& \frac{a_k v_S}{\sqrt{2}}+2 \lambda^2 v_S^2,
\eea
correspond to the eigenvectors
\bea
 h^0_1 &=& \frac{1}{\sqrt2} \( h_{dR}^0 + h_{uR}^0 \) ,\\
 h^0_2 &=& \frac{1}{\sqrt2 }\( h_{dR}^0 - h_{uR}^0 \).
\eea

Two distinct phenomenologically viable scenarios occur. Since we would like to identify one of the CP-even eigenstates
with the 125~GeV resonance hinted at by the LHC, it has to be made of doublets.
If $M_{h^0_1}^2 < M_{h^0_2}^2$ then, in the notation of the MSSM,  $h = h^0_1$, $H = h^0_2,$
and the Higgs mixing angle is given by $\a=-\pi/4=\b-\pi/2$. In that case $H$ does not have any direct tree level coupling to $WW$ and $ZZ$
that explains why the LHC does not see presently any other resonance, but the lightest one at 125~GeV.
On the other hand, if $M_{h^0_1}^2 > M_{h^0_2}^2$ then $h = h^0_2$, $H = h^0_1,$ and
the Higgs mixing angle is $\a=\pi/4=\b. $ In this case the LHC observed the second heaviest CP-even state because the couplings of the lightest one
 to fermions and to gauge bosons are strongly suppressed. Discovering such a light ``sterile" Higgs boson is very difficult at the LHC.

 The CP-odd Higgs boson  mass matrix in the basis $(h_{dI}^0,h_{uI}^0,s_I)$ is given by
\be {M'}_P^2 = \left(
\begin{array}{ccc}
 \frac{v_S^2 \l^2}{2} & \frac{v_S^2 \l^2}{2} & -\frac{v
   v_S \l^2}{\sqrt{2}} \\
 \frac{v_S^2 \l^2}{2} & \frac{v_S^2 \l^2}{2} & -\frac{v
   v_S \l^2}{\sqrt{2}} \\
 -\frac{v v_S \l^2}{\sqrt{2}} & -\frac{v v_S \l^2}{\sqrt{2}}
   & v^2 \lambda^2-\frac{3 a_k v_S}{\sqrt{2}}
\end{array}
\right), \ee
and the corresponding eigenvalues are
\bea
 M_{G^0}^2 &=& 0, \\
 M_{A^0_1}^2 &=& \frac{1}{4} \left(2 \l^2  \left(v^2+v_S^2\right)-3 \sqrt{2} a_kv_S
                 -\sqrt{\left(3 \sqrt{2} a_kv_S-2 \l^2    \left(v^2+v_S^2\right)\right)^2+24 \sqrt{2} a_k\l^2    v_S^3} \right), \quad \\
 M_{A^0_2}^2 &=& \frac{1}{4} \left(2 \l^2  \left(v^2+v_S^2\right)-3 \sqrt{2} a_kv_S
                 +\sqrt{\left(3 \sqrt{2} a_kv_S-2 \l^2    \left(v^2+v_S^2\right)\right)^2+24 \sqrt{2} a_k\l^2    v_S^3} \right).
\eea
While in (\ref{eq:MSnomix}) we cancelled the
singlet-doublet mixing in the CP-even sector by a particular choice of parameters, such a mixing still occurs in the CP-odd sector.

Finally the charged Higgs mass matrix in the basis $(h_u^+,
h_d^{-*} = h^+_d)$ is given by
\be
 {M'}_\pm^2 = \left( M_W^2+\frac{1}{2} \lambda^2 \left( 2 v_S^2 - v^2 \right)  \right)
\left(
\begin{array}{cc} 1 & 1 \\
                  1 & 1
\end{array}
\right), \ee
where $M_W^2=\frac{1}{4} g_2^2 v^2$. It contains one massless
Goldstone mode, and one massive eigenstate
\bea
 M_{H^\pm}^2 &=& M_W^2+\frac{1}{2} \lambda^2 \left( 2 v_S^2 - v^2 \right),
\label{MCHiggs}
\\
 H^+ &=& \frac{1}{\sqrt2} \(  h_{u}^+ + h_{d}^{-*} \).
\eea
The charged Higgs sector is identical to the MSSM one because $S$ is
electrically neutral. The matching can be easily done with the
substitutions
\bea
  \m &\to& \frac{1}{\sqrt2} \l v_S, \nn\\
   b &\to& \frac{1}{2} v_S^2.
 \label{eq:MSSMmapping}
\eea

Finally, we must ensure that all physical square masses are
positive, which is equivalent to checking that our solution is a
minimum of the potential. Moreover the constraint on $M_{H^\pm}^2$
implies that we are not breaking the $U(1)_{\rm em}$. Up to now we
only prevented the origin to be a minimum solution. So such a
requirement will impose further constraints on the free
parameters, that can be summarized as follows:
\be
 \text{sign}(a_k)=- \text{sign}(v_S), \quad  |a_k| < 2 \sqrt{2} \lambda^2 |v_S|,
\ee and one of the two following options
\begin{itemize}
 \item[a)] $v_S^2 > \frac{1}{2} v^2$,
 \item[b)] $\frac{1}{2} \l^2 \left(v^2-2 v_S^2\right) < M_W^2 .$
\end{itemize}

From now on we shall assume that the lightest CP even scalar is the one
coupled to the $W$'s, thus  $M_{h^0_1}^2 < M_{h^0_2}^2$. Moreover
we want also $M_{h^0_1}^2 > M_Z^2$.  This implies  that
the only available option is a). This fixes also the lightest CP-even Higgs couplings to gauge bosons to be
exactly as in the SM.
The lightest CP-even Higgs couples to the charged Higgs with
\bea
\lambda_{hH^+H^-}=2 c_W^2 - \frac{\lambda^2}{2}\frac{v^2}{M_Z^2},
\eea
where in our notation $c_W=\cos\theta_W$, $s_W=\sin\theta_W$ 
with $\theta_W$ the Weinberg angle, and
the coupling is normalized according to the conventions in 
\cite{Djouadi:2005gj}.
The first term is the MSSM contribution for $\a=-\pi/4=\b-\pi/2$, while the term in $\lambda$ is the NMSSM correction.

\begin{figure}[t]
\centerline{\epsfxsize = 0.5 \textwidth \epsffile{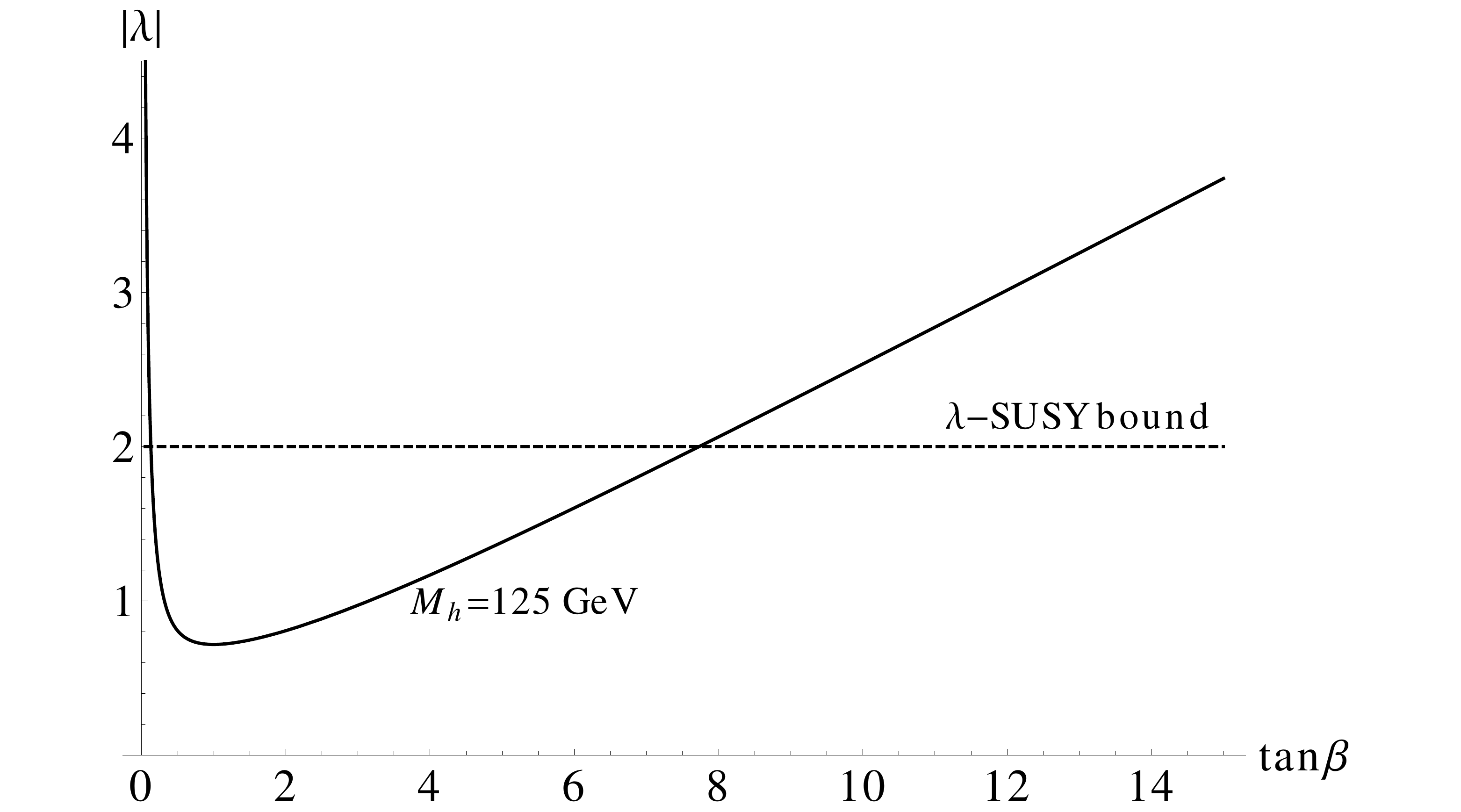} }
\caption{\it . Contour of $M_h= 125$~GeV in the $(\tan\b,|\l|)$ plane (solid line). The dashed line is the
$\l$-SUSY upper bound. \vspace*{0.5cm}} \label{hplot}
\end{figure}

We now consider the impact of relaxing the  $\tan\b=1$ condition in the present analysis.
We assume as usual that the scalar singlet is very heavy, and $k v_S \gg M_Z, a_k, a_\l$.
Within such approximation it is easy to derive \cite{nmssm} that  $\a \simeq \b-\pi/2$ still holds,
and $H$ is again essentially decoupled from $WW$ or $ZZ$. On the other hand, one has
\be
 M_h^2 \simeq M_Z^2 \cos^2(2 \beta )+\frac{1}{2} \lambda^2 v^2 \sin^2(2 \beta ) - \frac{\l^2}{k^2} v^2 \(\l - k \sin(2\beta) \)^2 \, .
\ee
Hence, the presence of the singlet can still give a negative contribution to the light Higgs boson mass.
To prevent the latter negative contribution, we generalize eq. (\ref{eq:nomixingSH2}), and assume
\be
 k = \l/\sin(2\beta)
\ee
so that
\be
 M_h^2 \simeq M_Z^2 \cos^2(2 \beta )+\frac{1}{2} \lambda^2 v^2 \sin^2(2 \beta ) \, .
\ee
Here, we are interested in values of $\l$ and $\tan\b$ that satisfy $M_h\approx 125$~GeV.
In Fig. \ref{hplot} we plot the corresponding two-dimentional curve of $|\l|$ versus $\tan\b$.
The black continuous line represents $M_h= 125$~GeV.
In the $\l$-SUSY theory \cite{Barbieri:2006bg}, $\l$ is increased so that the interaction becomes non-perturbative
below the unification scale. However, $\l$ should not exceed $\sim$ 2, otherwise non-perturbative physics
would appear below 10 TeV, spoiling our understanding of precision electroweak
data in the perturbative theory. The dashed line in Fig. \ref{hplot} represents the $\l$-SUSY upper bound. Then,
only low values of $\tan\b$ are allowed, in the range $\tan\b<8$.
In particular, $\tan\b=1$ corresponds  to the minimal $\l$ value ($\l \simeq 0.72$).

Finally,  the relevant quantities for the charged Higgs phenomenology in the general $\tan\b$ case are
\bea
 M_{H^\pm}^2 &=& \frac{\lambda^2 v_S^2}{ \sin^2(2 \beta )}+M_W^2-\frac{\lambda ^2 v^2}{2} \\
\lambda_{hH^+H^-} &=& \cos(2 \b) \sin(\a + \b) + 2 c_W^2 \sin( \b - \a) - \frac{\lambda^2}{2}\frac{v^2}{M_Z^2} \cos(\a + \b) \sin(2 \b)
\label{eq:hHpHm}
\, .
\eea

\subsection{Neutralinos and charginos}
The soft SUSY breaking gaugino mass terms in the Lagrangian read
\be
 {\cal L}  = - \frac{1}{2} \( M_1 \l_1 \l_1 +  M_2 \l_2^i
\l_2^i  +  M_3 \l_3^a \l_3^a \).
\ee
In the basis $\psi^0 = (\l_1 , \l_2^3, \tilde h_d^0, \tilde h_u^0,
\tilde s)$, the resulting mass terms in the Lagrangian read
\be {\cal L} = - \frac{1}{2} (\psi^0)^T {\cal M}_0 (\psi^0) +
\mathrm{h.c.} \ee where \be
 M_0 =
\left( \begin{array}{ccccc}
   M_1 &   0   & -M_Z s_W \cos\b  &  M_Z s_W \sin\b      & 0 \\
 \dots & M_2   &  M_Z c_W \cos\b & -M_Z c_W \sin\b       & 0 \\
 \dots & \dots &        0          & -\frac{1}{\sqrt2} \l v_S & -\frac{\l v}{\sqrt2} \sin\b \\
 \dots & \dots &      \dots        &   0                     & -\frac{\l v}{\sqrt2} \cos\b \\
 \dots & \dots &      \dots        & \dots                   & \sqrt{2} k v_S
\end{array} \right).
\ee
Because of supersymmetry and electric charge conservation the
chargino sector is the same as the MSSM up to the substitutions
(\ref{eq:MSSMmapping}), in the gauge-eigenstate basis $\psi^\pm =
(\tilde W^+,\, \tilde H_u^+,\,\tilde W^- ,\, \tilde H_d^- )$ the
chargino mass terms in the Lagrangian are
\be {\cal L} _{\mbox{chargino mass}} = -\half (\psi^\pm)^T {\bf
M}_{\tilde C} \psi^\pm + h.c. \ee
where, in $2\times 2$ block form,
\be M_{\tilde C} = \( \begin{array}{cc}
{\bf 0}   & {\bf X}^T\\
  {\bf X} & {\bf 0}
\end{array}
\) ,\ee
with
\be {\bf X} = \( \begin{array}{cc}
 M_2 &  \sqrt2 M_W \sin\b \\
   \sqrt2 M_W \cos\b & \frac{1}{\sqrt2} \l v_S
\end{array} \).
\label{eq:charginomassmatrix} \ee

\subsection{Radiative Higgs boson decays}

The model we have chosen to work with leaves the FP Higgs boson decays to $WW^*$ and $ZZ^*$ final states
at tree level unaffected compared to the FP SM predictions. Although it is easy in  this framework to decrease the coupling
at tree level by choosing a different $\tan\beta$ and the Higgs mixing parameter $\alpha,$ in the FP Higgs scenario
this will also suppress the induced $\gamma\gamma$ rate because the latter is dominated by the $W$-boson loop.
Because fermiophobia by itself is able to explain the observed deficit in $WW^*$ channel~\cite{Gabrielli:2012yz},
our choice is motivated by a maximized  $\gamma\gamma$ rate. Therefore the deviations from the FP SM predictions
may happen only due to extra particles in the loop. Because in the FP Higgs scenario the flavour physics constraints on charged Higgs masses are largely
removed and chargino could be light, those particles can be as light
as their present lower bounds from LEP II.

The free parameters at the EW scale which are
relevant for our analysis are the following: $\tan\beta$,
the gaugino masses $M_1$ and $M_2$, the $\mu$ term given by
$\mu\equiv \lambda v_S/\sqrt{2}$, the sign($\mu M_1)$
and sign($\mu M_2)$. In the present model, the mass of the charged Higgs
is fixed once the value of $\mu$ is given, see Eq.(\ref{MCHiggs}).
Moreover, we have chosen the convention of keeping $M_{2}$
positive, and allowing sign($\mu)$ to vary.
We have set $M_h=125$~GeV, $M_1=100$~GeV and $\sqrt{s}=7$~TeV.

Then, $\tan\beta$,  $|\m|$, sign($\mu)$ and $M_2$ are free parameters.
We chose to re-express  $|\mu|$ and $M_2$ as functions of two physical mass parameters:
the charged Higgs mass ($M_{H^+}$) and the lightest chargino mass ($M_{\chi^+_L}$), as follows
\bea
 |\m| &=& \frac{\sqrt{M_h^2+\left(M_{H^+}^2-M_W^2\right)\sin^2(2 \beta) -M_Z^2 \cos ^2(2 \beta) }}{\sqrt{2}} \\
 M_2 &=& \frac{\pm \sqrt{2} M_{\chi^+_L} \sqrt{4 M_W^2 \left(\mu ^2-M_{\chi^+_L}^2\right)+2 \left(M_{\chi^+_L}^2-\mu^2\right)^2+M_W^4 \(1 - \cos 4\beta \)}
         -2 \mu  M_W^2 \sin 2 \beta }{2 (M_{\chi^+_L}^2-\mu^2 )} \, .\nn\\ \label{eq:M2}
\eea
There are two different values of the gaugino mass $M_2$, corresponding to the same lightest chargino mass.
For convention  $M_{2}>0$ and, for each  sign($\mu$) and $|\mu|$ value, only  one of the two solutions is allowed.
Finally, we recall that the $\l$ parameter is determined by $M_h$ and $\b$ as
\be
 |\l| = \frac{\sqrt{2} \sqrt{M_h^2-M_Z^2 \cos ^2(2 \beta )}}{v \sin(2\beta)}\, .
\ee
In Table~1,  we give some numerical values of the input parameters and the corresponding {\it derived} fundamental parameters, for $M_h=125$ GeV and $M_{H^+}=400$ GeV.
  \begin{table}[ht]
  \centering
  \begin{tabular}[h]{|c|c|c|c|c|c|c|c|}
 \hline \multicolumn{3}{|c|}{Input parameters} & \multicolumn{3}{|c|}{Derived parameters} \\ \hline \hline
sign($\m$) & $\tan\b$  & $M_{\chi^+_L}$ (GeV) &  $|\l|$ & $|\mu|$ (GeV) & $M_2$ (GeV) \\ \hline
+ & 1 &  200 &  0.72 & 290.8 & 271.2 \\   \hline
+ & 5 &  200 &  1.40 & 125. & 125.8 \\    \hline
- & 1 &  200 &  0.72 & 290.8 & 186.8 \\   \hline
- & 5 &  200 &  1.40 & 125. & 151.3 \\    \hline
+ & 1 &  400 &  0.72 & 290.8 & 340.8 \\   \hline
+ & 5 &  400 &  1.40 & 125. & 379.6 \\    \hline
- & 1 &  400 &  0.72 & 290.8 & 390.6 \\   \hline
- & 5 &  400 &  1.40 & 125. & 383.9 \\    \hline
  \end{tabular}
  \caption{Numerical values of the input parameters and the corresponding derived fundamental parameters for $M_h=125$ GeV and $M_{H^+}=400$ GeV.}
  \end{table}

The corresponding decay widths for the radiative induced decays
$h\to \gamma \gamma$ and $h\to Z \gamma$
of the lightest CP even Higgs boson $h$, in the framework of pure
FP NMSSM model, are reported in Appendix.

In order to avoid a large tree-level Higgs decay into an invisible sector~\cite{Raidal:2011xk},
that would destroy the potential enhancement of the
Higgs decay into $\gamma \gamma$~\cite{Giardino:2012ww}, we will
require that the mass of the lightest neutralino ($M_{\chi^0_L}$),
which is the lightest
supersymmetric state in our scenario, is heavier than half of the Higgs mass.
Then, due to $R$-parity, all other Higgs decays into two generic neutralino
states $h\to \chi^0_i \chi^{ 0}_j$,
including the case in which one or both are virtual states,
will automatically vanish. In addition, we also require
the lower bound on the chargino mass to be 90 GeV.

\begin{figure}[t]
\centerline{\epsfxsize = 0.5 \textwidth \epsffile{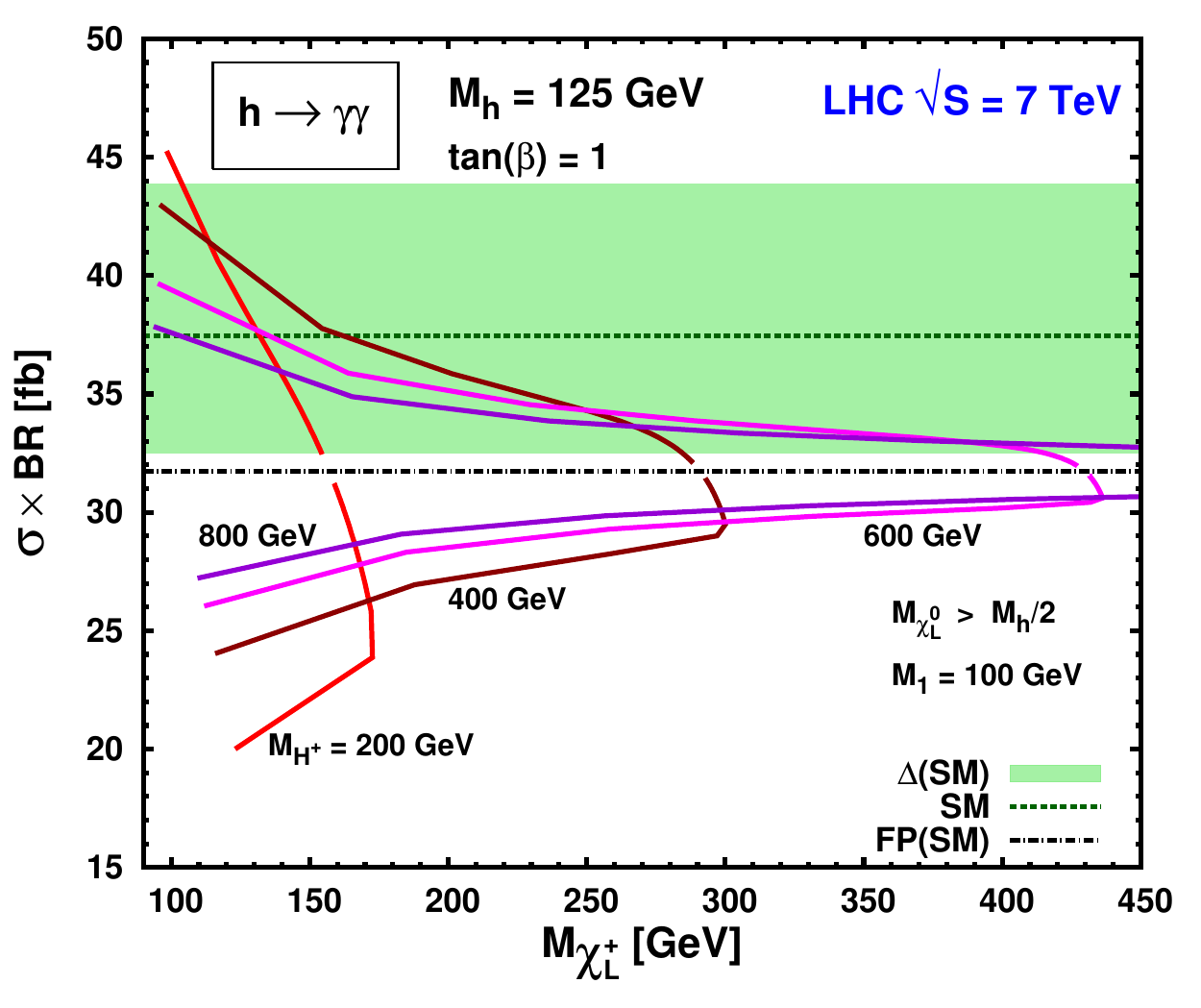}
\hfill \epsfxsize = 0.5 \textwidth \epsffile{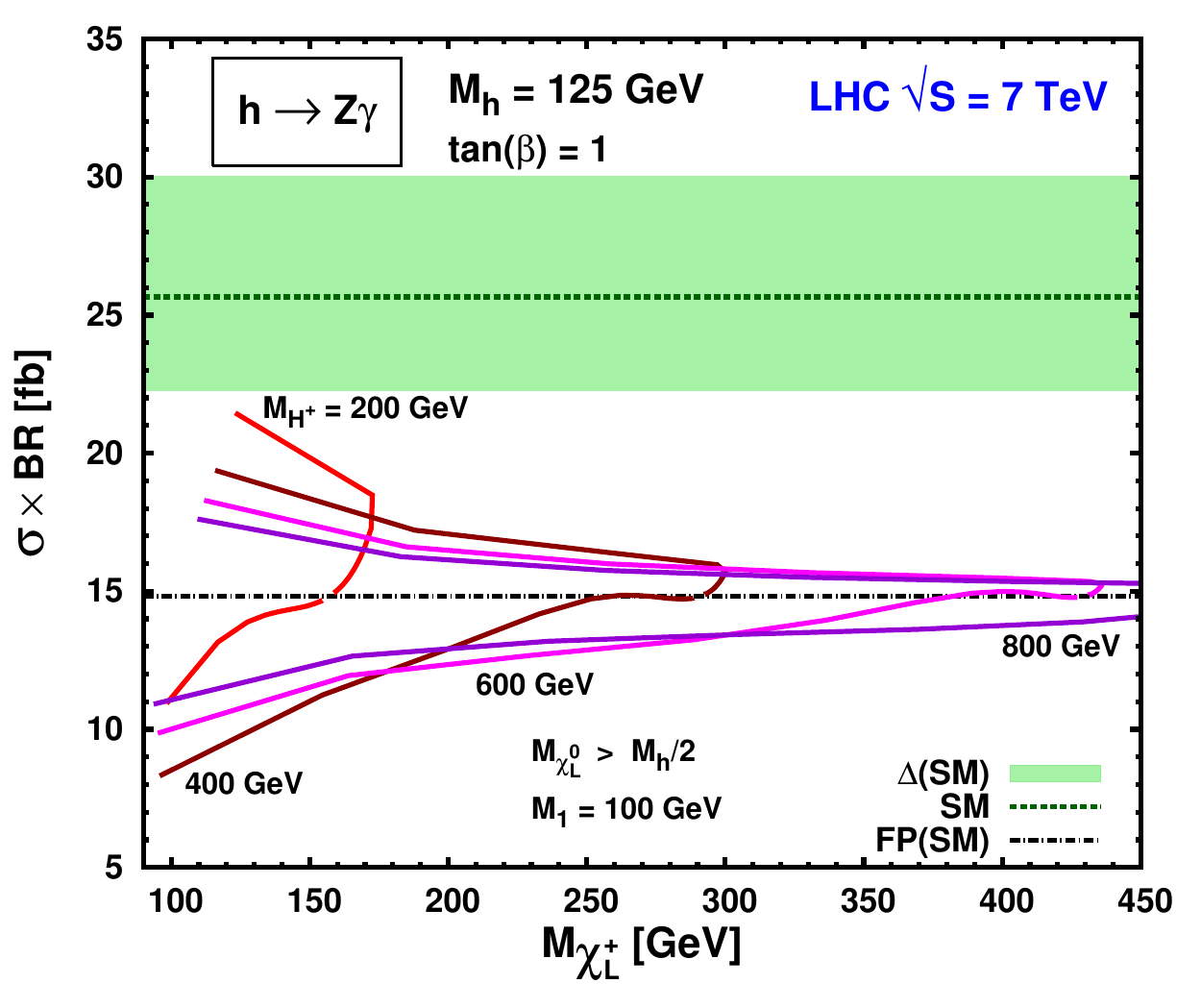}
}
\caption{\it
Radiatively induced signal rates of 125~{\rm GeV} FP NMSSM  Higgs decays $h\to \gamma\gamma$ (left)  and $h\to Z\gamma$ (right) at the 7~{\rm TeV} LHC
as functions of the lightest chargino $\chi^+_L$ mass ($M_{\chi^+_L}$)
for several charged Higgs boson $H^+$ masses ($M_{H^+}$) as indicated in figures and for $\tan\beta=1$. The SM central value prediction (dashed line)
together with $1\sigma$ uncertainty band $\Delta(SM)$
and the FP SM Higgs prediction are also shown. The lines above (below)  the FP SM line correspond to $sign(M_2\mu)>0$ ($sign(M_2 \mu)<0$) for $h\to \gamma \gamma$ and
to $sign(M_2\mu)<0$ ($sign(M_2 \mu)>0$) for $h\to Z \gamma$.
\vspace*{0.5cm}}
\label{fig1}
\end{figure}
\begin{figure}[htb]
\centerline{\epsfxsize = 0.5 \textwidth \epsffile{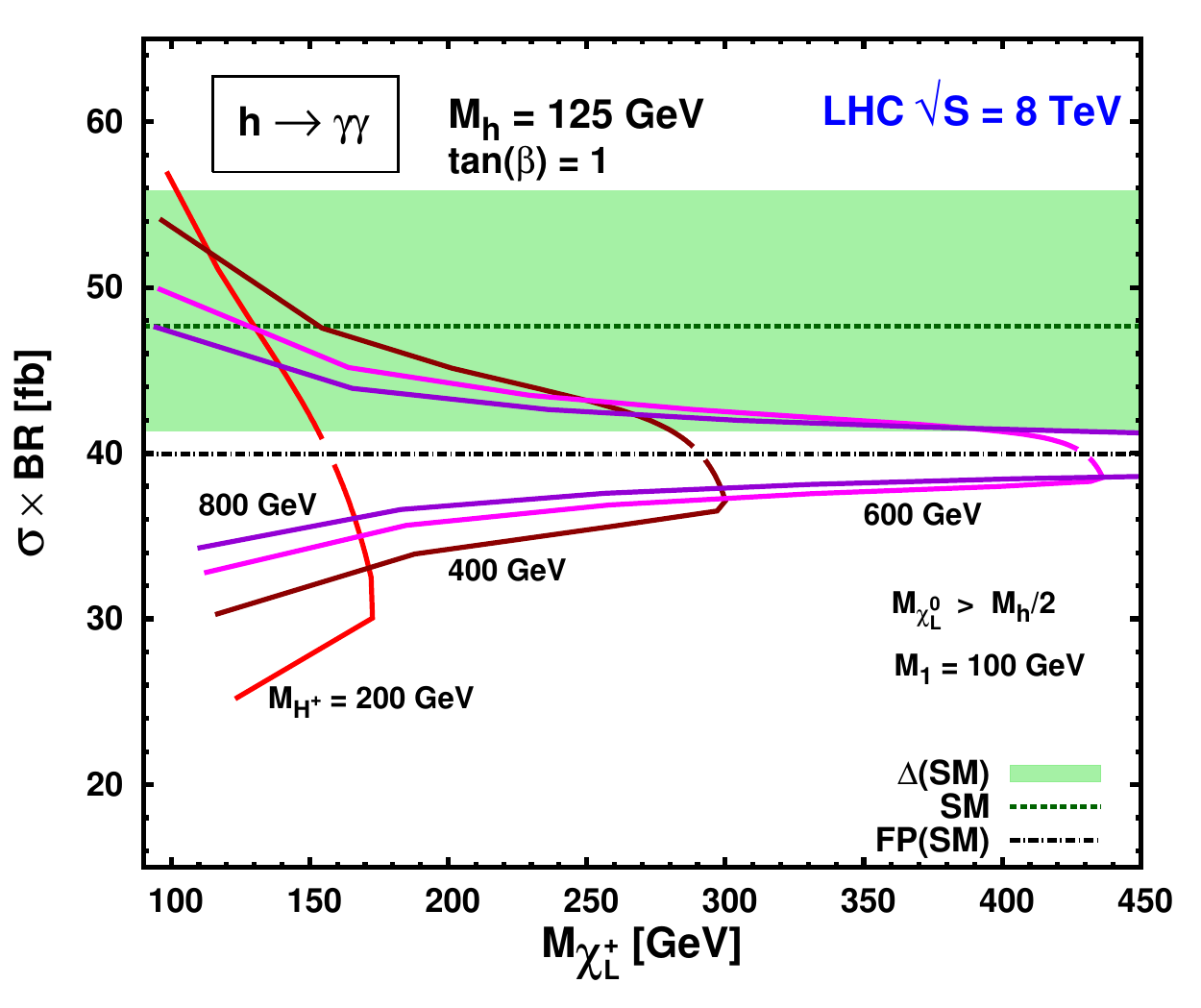}
\hfill \epsfxsize = 0.5\textwidth \epsffile{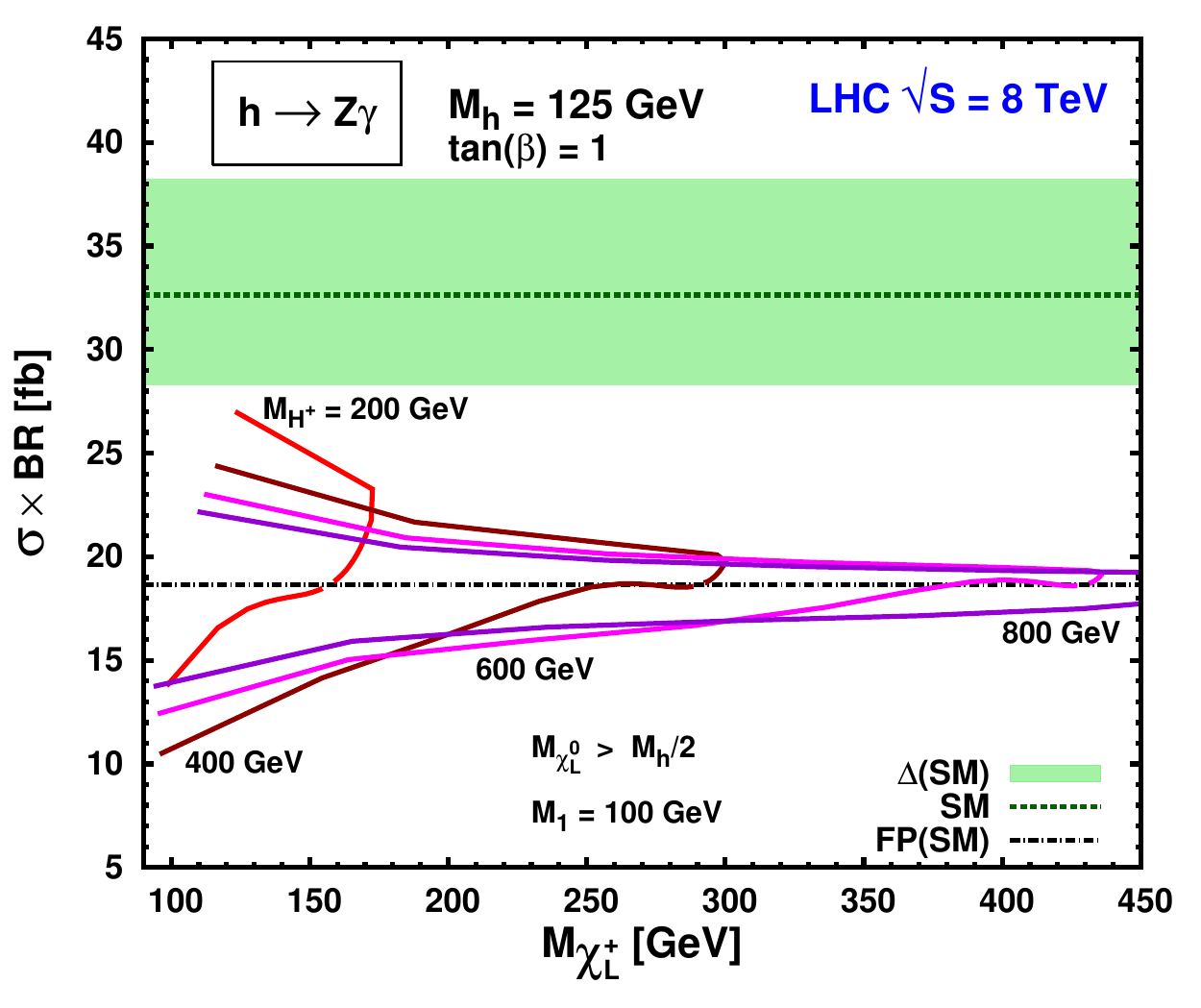}
}
\caption{\it
The same as in Fig.~\ref{fig1} but for a $M_h=125$~{\rm GeV} FP Higgs boson at the $8$~{\rm TeV} LHC.
\vspace*{0.5cm}}
\label{fig2}
\end{figure}

\begin{figure}[htb]
\centerline{\epsfxsize = 0.5\textwidth \epsffile{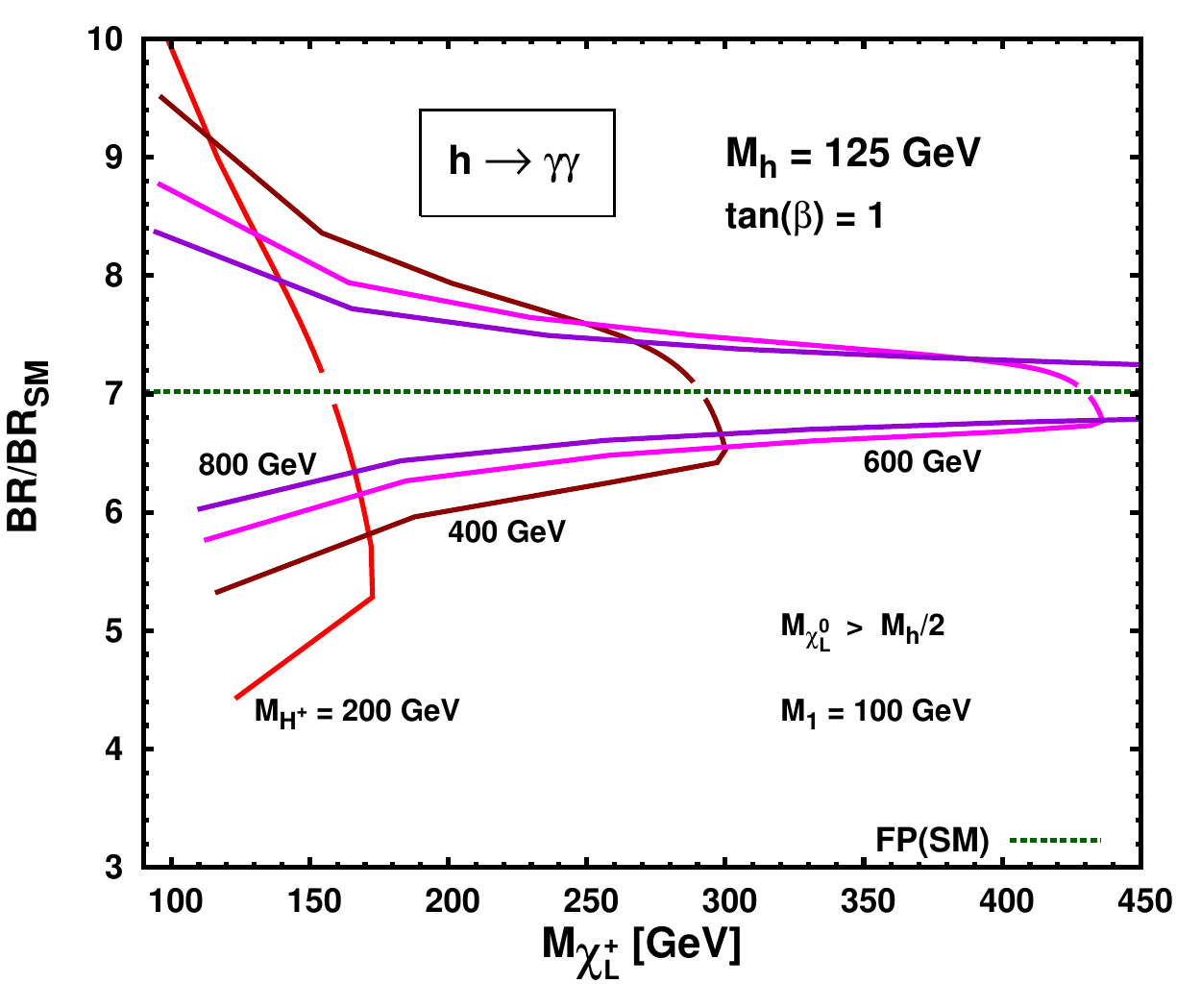}
\hfill \epsfxsize = 0.5\textwidth \epsffile{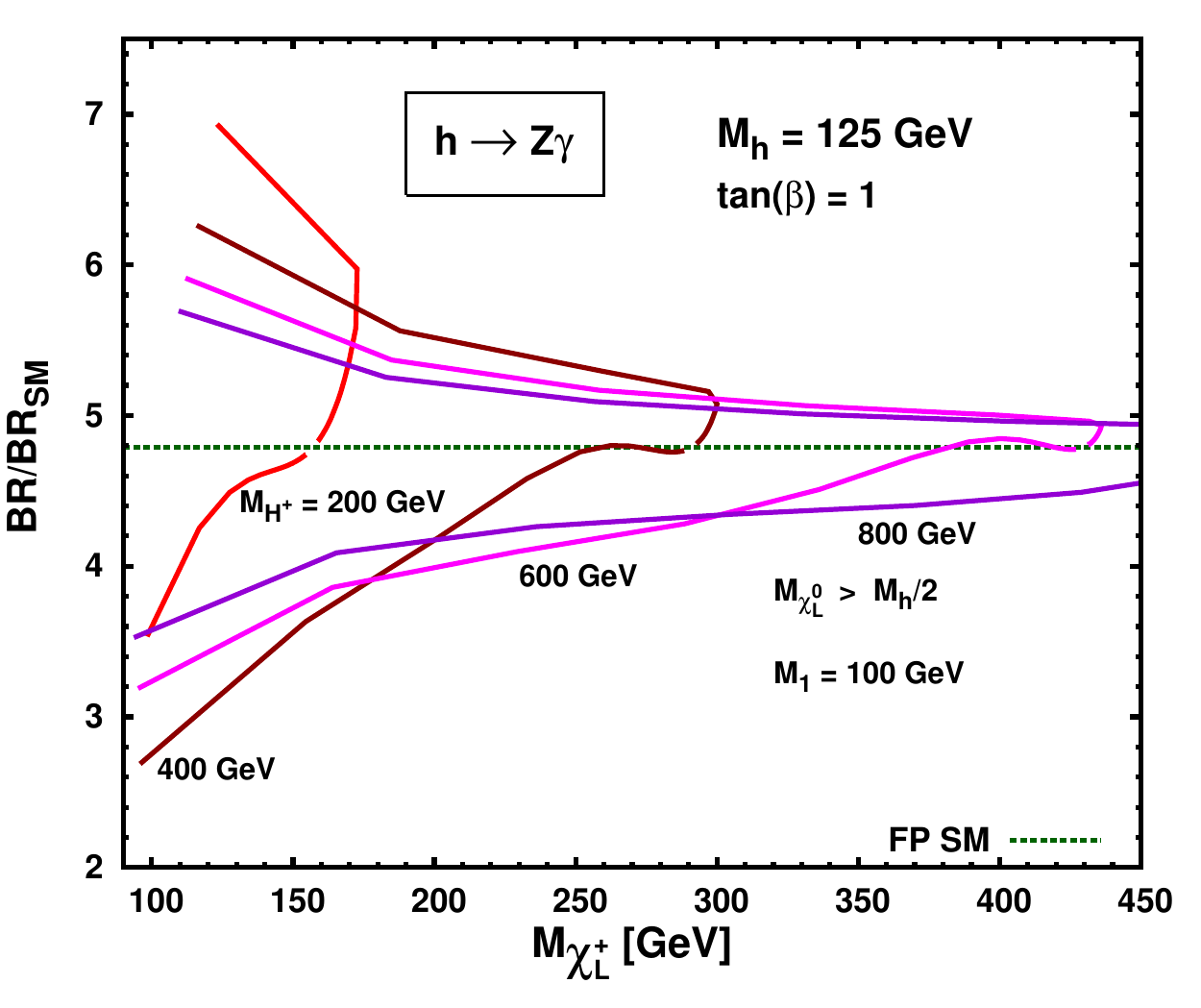}
}
\caption{\it
Dependence of the FP NMSSM  Higgs boson branching fraction over the SM value as a function of the lightest chargino mass $M_{\chi^+}$ for $M_h=125$~{\rm GeV},
for different values of the charged Higgs boson mass $M_H^+$
and for $\tan\beta=1$. The dashed line corresponds to the FP SM scenario.
The lines above (below)  the FP SM line correspond to $sign(M_2\mu)>0$ ($sign(M_2 \mu)<0$) for $h\to \gamma \gamma$ and
to $sign(M_2\mu)<0$ ($sign(M_2 \mu)>0$) for $h\to Z \gamma$.
\vspace*{0.5cm}}
\label{fig3}
\end{figure}

Taking into account the results obtained so far, we have computed the $h\to \gamma\gamma$  and $h\to Z\gamma$
signal rates for the FP Higgs boson in the FP NMSSM. We present our results in Fig.~\ref{fig1} where we plot  the
signal rates of those processes as functions of the lightest chargino mass
$M_{\chi^+_L}$ for different charged Higgs boson masses
as indicated in the figure.
The SM predictions together with their uncertainties and the FP SM predictions are also presented. The $1\sigma$ (green) band corresponds to the
theoretical uncertainty on the SM production cross section by
gluon-gluon fusion. We have not included the uncertainty band on
the SM FP line since the corresponding theoretical uncertainty
due to the VBF cross section is quite small and can be neglected
in this context.

As in the MSSM case, the dominant contribution to SUSY contribution to
$h\to \gamma \gamma$ and $h\to Z \gamma$ amplitudes
comes from the charginos loop.
From Figs.~\ref{fig1}, \ref{fig2}, we can see that for fixed
chargino mass there are always two
solutions for the one-loop SUSY amplitudes corresponding to
$h\to \gamma \gamma$ and $h\to Z \gamma$ decays.
This can be understood as follows.
For values of $M_{\chi^+_L}$ below the interesection point with
the FP(SM) line, the double solution is mainly due to the sign of $\mu$ that
controls the relative sign of the SUSY amplitude with respect to the SM one.
The lines above (below)  the FP SM line correspond to $sign(M_2\mu)>0$ ($sign(M_2 \mu)<0$) for $h\to \gamma \gamma$ and to $sign(M_2\mu)<0$ ($sign(M_2 \mu)>0$) for $h\to Z \gamma$. The dependence of the curves by $sign(M_2\mu)$ can be
understood from Eqs.(\ref{Gammagg},\ref{GammaZg}) in appendix.

However, for values  of $M_{\chi^+_L}$ above the intersection point,
the sign of $\mu$ is fixed and the double solution corresponds
to the fact that in the loop run two non-degenerates values of heavy chargino
states at fixed $M_{\chi_L^+}$.
These two values of heavy chargino mass, for a fixed light chargino mass, correspond to the two different solutions for the $M_2$ parameter in eq. (\ref{eq:M2}).
The kink point corresponds to
the case where the two solutions for the heavy chargino masses coincide.
As we can see, there is a non-decoupling effect of the SUSY contribution to
the loop $\gamma \gamma$ and  $Z \gamma$ decay amplitudes
in correspondance of the maximum value for the lighest chargino mass.

As seen in the Figs.~\ref{fig1},~\ref{fig2}, the FP NMSSM signal rates can be
both bigger or smaller than the FP SM predictions. For very light sparticles the
total rate in $\gamma\gamma$ channel can even exceed the SM prediction. On the other hand, the present fits indicate that the
LHC observes fewer $\gamma\gamma$ events than predicted by the pure FP SM~\cite{Giardino:2012ww}.
This result can be easily explained in the
FP NMSSM since also rate reductions of  as much as 50\% are possible for the chosen parameters.
The absence of points in the half-plane above (below) the FP(SM) line
for the curve corresponding to $M_{H^+}=200$ GeV in the case of
$h\to \gamma \gamma$ ($h\to Z\gamma$), is due to the lighest neutralino mass
constraint $M_{\chi_L^0} > M_h/2$ and depends on our choice for $M_1=100$ GeV.

At the 8~TeV LHC the predictions are qualitatively the same but  numerically different. We
present the rates for 8~TeV LHC in Fig.~\ref{fig2} for the same model parameters as in Fig.~\ref{fig1}.

If the top Yukawa coupling of the Higgs boson is not exactly vanishing, also gluon-gluon fusion production process
will contribute to the Higgs production. In that case it is important to know
our predictions for the FP Higgs branching
fractions in our scenario. In Fig.~\ref{fig3} we plot  the deviation of
FP NMSSM Higgs boson branching fractions from the SM prediction
for the previously specified parameters. The qualitative behaviour of
branching fractions is the same as in previous figures,
explaining our results.

\begin{figure}[t]
\centerline{\epsfxsize = 0.5 \textwidth \epsffile{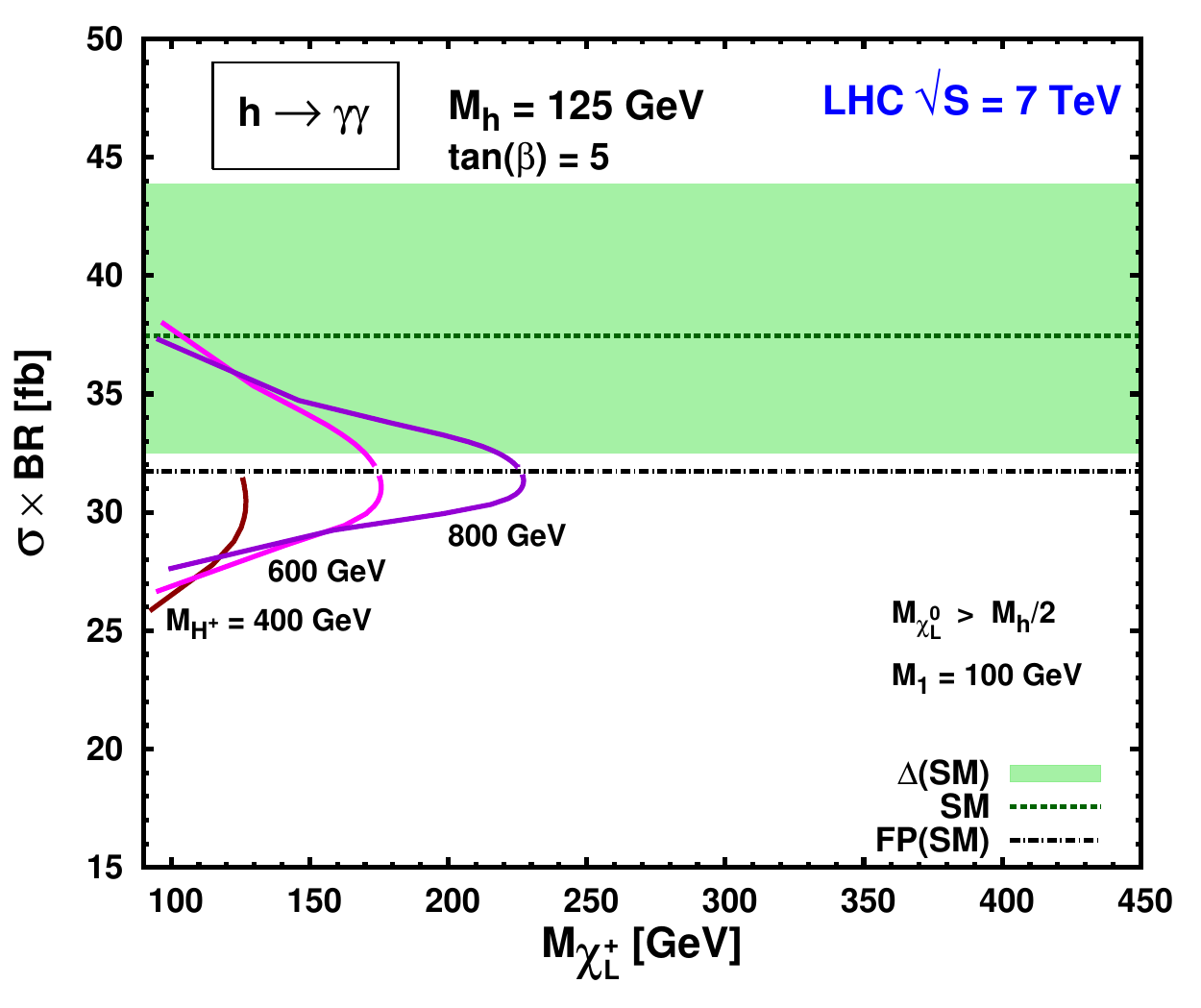}
\hfill \epsfxsize = 0.5 \textwidth \epsffile{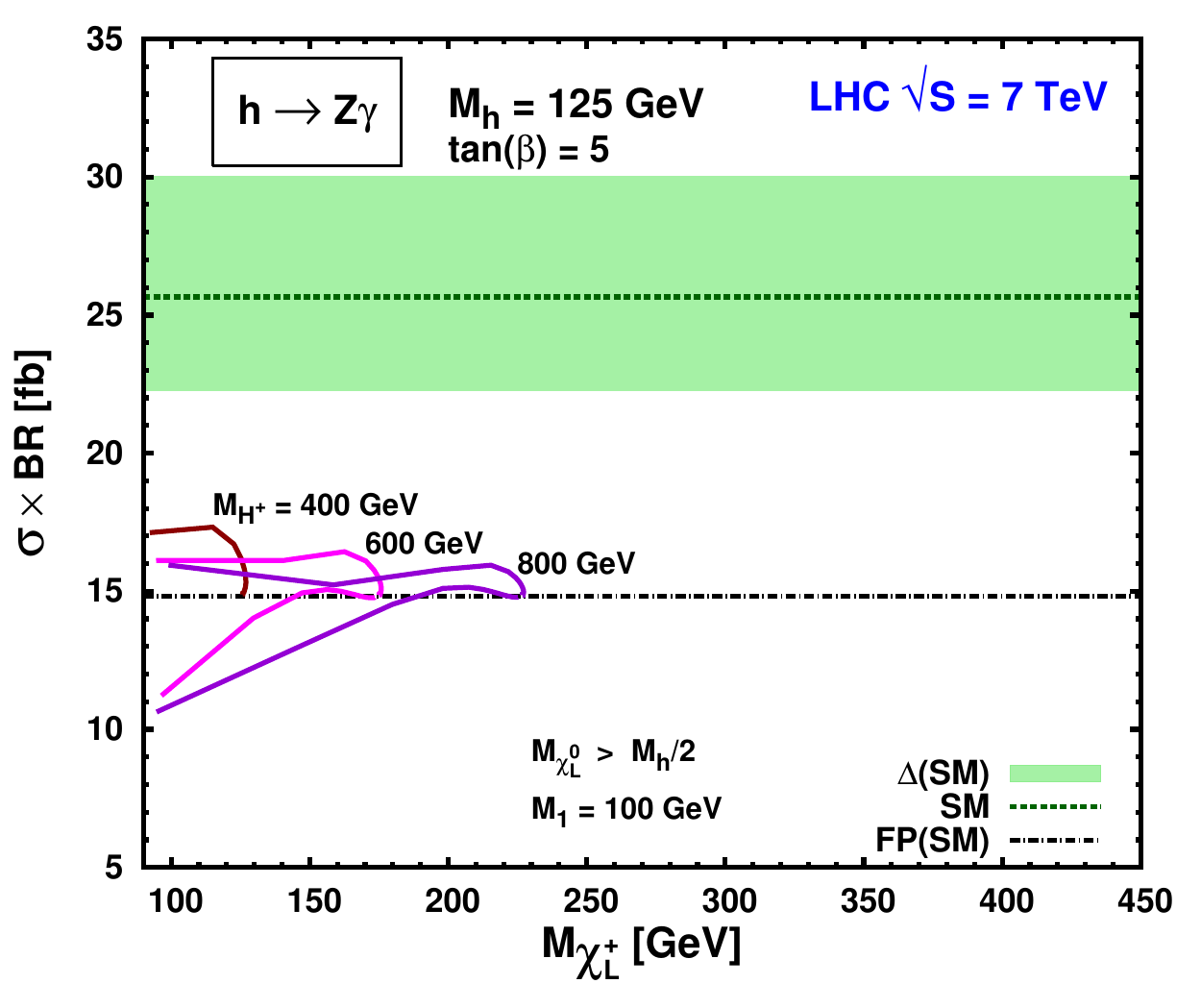}
}
\caption{\it The same as in Fig.~\ref{fig1}, but for $\tan\beta=5$.
\vspace*{0.5cm}}
\label{fig4}
\end{figure}

\begin{figure}[t]
\centerline{\epsfxsize = 0.5 \textwidth \epsffile{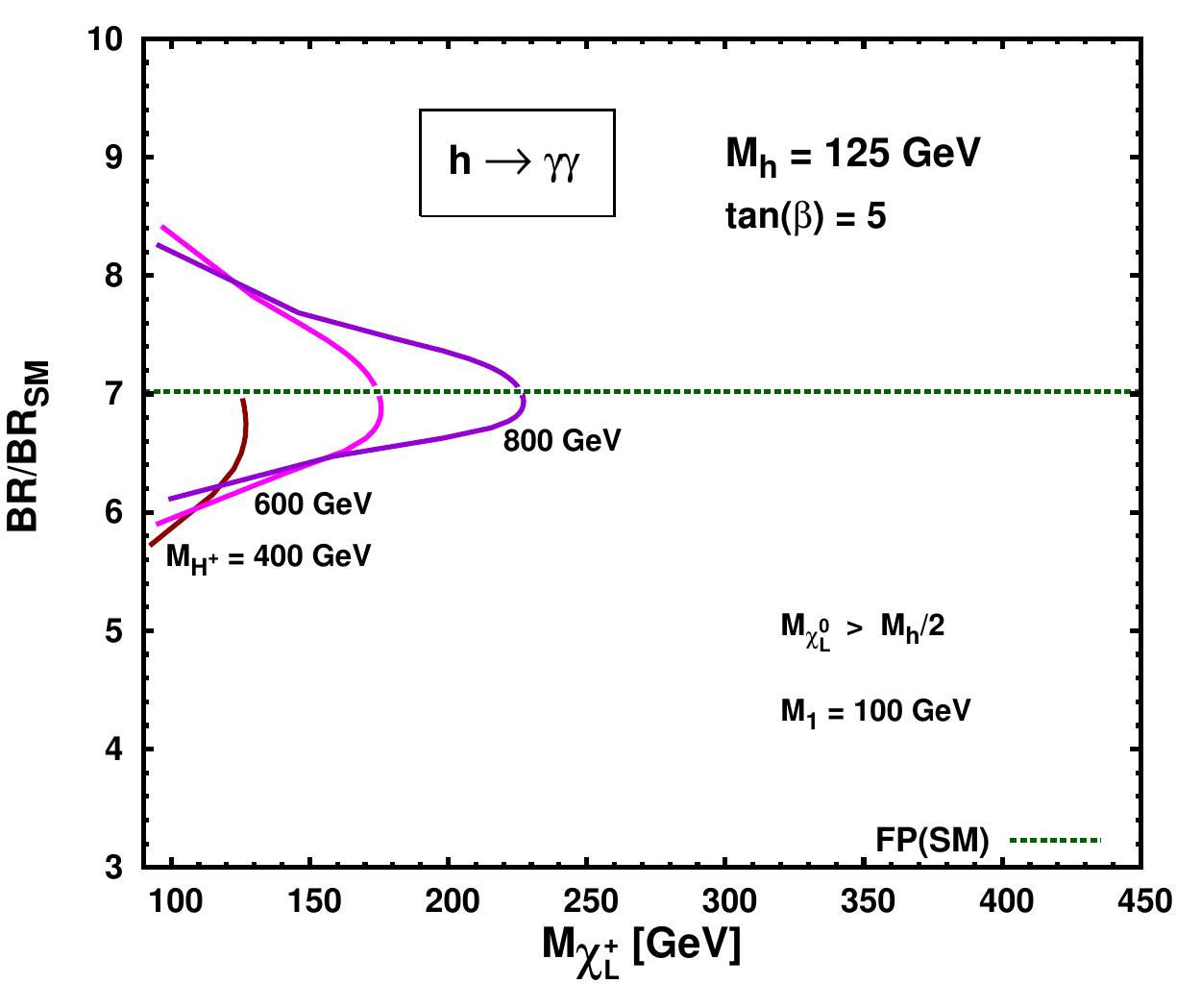}
\hfill \epsfxsize = 0.5 \textwidth \epsffile{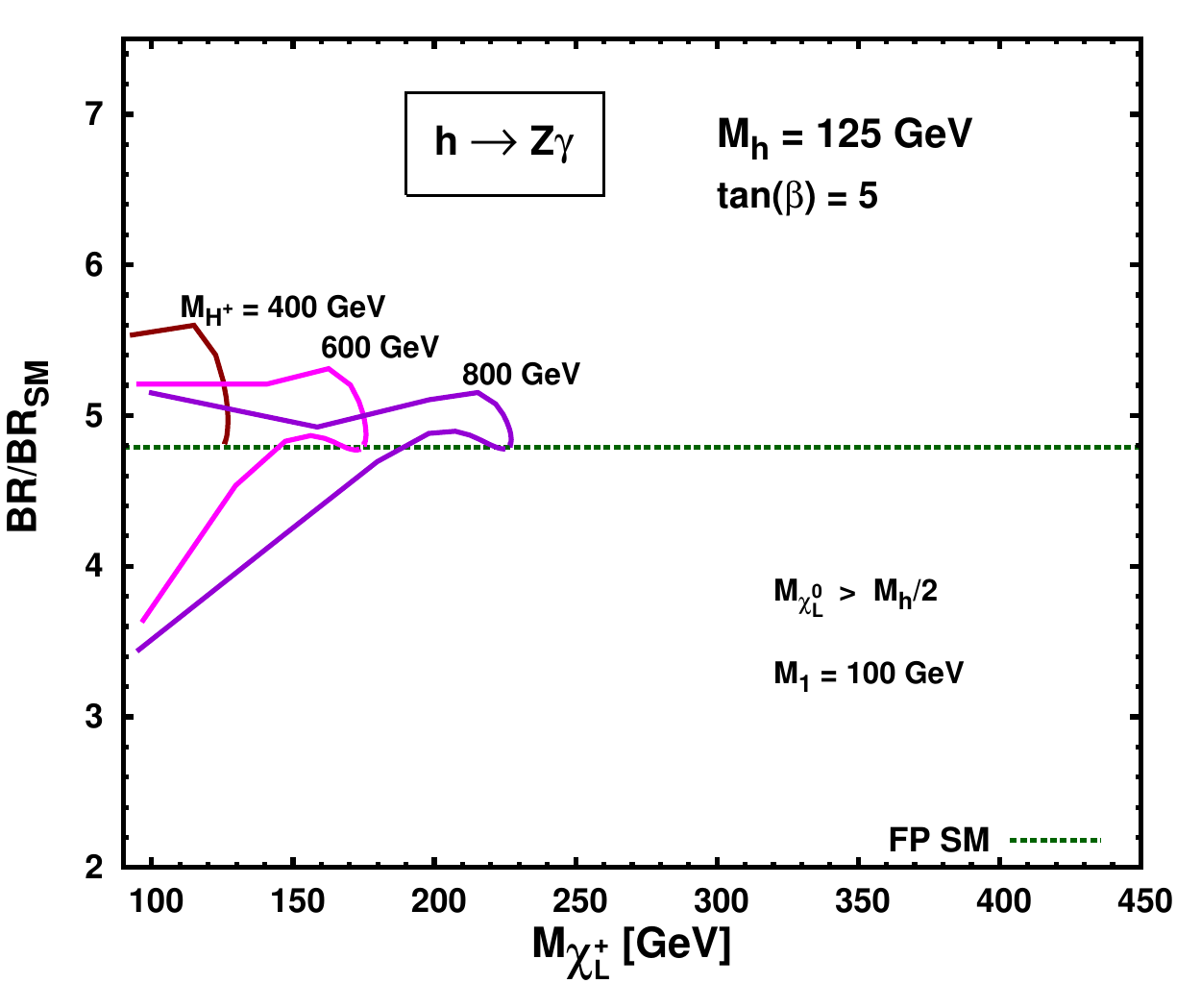}
}
\caption{\it The same as in Fig.~\ref{fig3}, but for $\tan\beta=5$.
\vspace*{0.5cm}}
\label{fig5}
\end{figure}

Finally, in Figs.~\ref{fig4} and  ~\ref{fig5} we present
the same plots as in Figs.~\ref{fig1} and ~\ref{fig3}, respectively,
for $\tan\beta =5$.
We set this value as an intermediate point between $\tan\beta=1$ and the
$\lambda$-SUSY upper bound $\tan\beta < 8$  at $M_h= 125$ GeV
(see  Fig.~\ref{hplot}).
By increasing
$\tan\beta$, the deviations from the SM FP Higgs predictions for
production rates  and BR's for
$h\to \gamma \gamma$ and $h\to Z\gamma$ are quite decreased. The largest effect
is indeed achieved for $\tan \beta=1$. On the other hand, at $\tan\beta=5$, the largest SUSY
contribution  is obtained for a charged Higgs mass
$M_{H^+} \sim 400$ GeV and a light chargino mass $M_{\chi^+} < 150$ GeV.
The curves corresponding to $M_{H^+}=200$ are not present in
Figs.~\ref{fig4} and  ~\ref{fig5},  not being allowed for $\tan\beta =5$, because of
the constraint  $M_{\chi^0} > M_h/2$.

\section{Discussion}

Apart from the Higgs boson phenomenology at the LHC discussed in the
previous section,
the FP NMSSM scenario has other important implications for SUSY phenomenology.
As we have emphasized, the $b\to s\gamma$ and $B_s\to \mu\mu$ constraints on the charged Higgs mass are absent in this model.
Thus the charged Higgs boson can be light and kinematically accessible at
the LHC in the process $pp\to H^+H^-.$ The same may apply to other
possible scalar and pseudo-scalar final states. While we have chosen to work with a particular model in which the Higgs boson is
exactly SM-like, in general also other final states are possible, allowing to study the model parameters. However, because they are
fermiophobic, their search strategy must be revised.

If neutralinos and charginos $\chi_i$ are light, the dominant decay modes of all heavy Higgs bosons $H_i$ could be into two  $\chi_i.$
In particular, the tree level decays of $s_R$ are induced  by the $\l$ coupling.
If the decay channels to sparticles  are kinematically forbidden,
the heaviest among $A^0_1, A^0_2, H, H^\pm$ will have tree level
decays into the lighter ones and to (real or virtual) $W$'s or $Z$'s.
Then the lightest of them, since it cannot decay into SM fermions
because of fermiophobia, will have SUSY induced radiative decays giving in
the final state the SM gauge bosons and fermions. We stress that,
because of the decoupling induced by our values of
$\a$ and $\b$, there are no trilinear vertices involving only one
$h$ and one of the scalars among $A^0_1, A^0_2, H, H^\pm$.
Thus all the decays of heavier Higgs bosons are characterized either by large invisible branching fraction or
 multi-particle final states. Those decay signatures can easily be missed in
present LHC searches explaining the absence of another Higgs-like resonance at higher masses.

The second most relevant phenomenological implication of our framework concerns direct dark matter searches.
In the MSSM the spin-independent dark matter scattering off nuclei is dominated by tree level Higgs boson exchange.
In the FP Higgs case this process is suppressed. The dominant dark matter scattering process is through $WW$ exchange at one
loop level. This implies that the scattering cross section is suppressed by additional loop factor compared with the MSSM
expectations. Scattering due to $W$-loops is too weak~\cite{Hisano:2010fy} to be observed in the present stage of XENON100~\cite{Aprile:2011hi}.

Arguably the   biggest drawback of our scenario is the absence of explanation for  the third generation fermion masses.
However, models of composite Higgs boson that explain naturalness of the electroweak symmetry breaking with new
strong dynamics at 2--3~TeV scale do predict non-standard Higgs boson coupling to fermions~\cite{Heff}. Fermiophobia may be
a feature of this framework. Supersymmetrizing the theory will stabilize the radiatively generated Yukawa couplings against
new physics at high scales. While in non-SUSY case one expects large radiative corrections to Yukawa couplings
proportional to $\log (M_h/\Lambda),$ where $\Lambda$ is the unknown scale of new physics, in the SUSY version of FP Higgs
those corrections will be at most of order  $\log (M_h/M_{\rm SUSY}),$ hence stabilizing the theory.

\section{Conclusions}

If there is a signal of a fermiophobic, or partially fermiophobic,
Higgs boson with mass $M_h=125$ GeV, the fundamental idea of
supersymmetry, as it is implemented in the MSSM, is in trouble and
must be revised.  In particular, we have shown that the MSSM with vanishing
or strongly suppressed Yukawa couplings is ruled out,
independently of the particular supersymmetry breaking mechanism.
Indeed, due to the absence of Yukawa couplings the usual
(large) logarithmic corrections to the Higgs mass, induced by the scalar particles running
in the loops,  are absent and the upper bound on the Higgs mass
is very close the $M_Z$ mass.

In order to rescue supersymmetry, we show that a viable model beyond MSSM could
be the NMSSM,
where the absence of tree-level Yukawa couplings in the superpotential
is guaranteed by the addition of a  $Z_3$ discrete symmetry.
The most relevant aspects of this scenario is that the
SUSY naturalness criteria are automatically relaxed by a factor $N_c y_t^4/g^4 \sim 25$,
solving the little hierarchy problem and allowing sparticle masses to be naturally
of order 2--3 TeV. Moreover, the usual flavor and CP problems are  relaxed
partly because of the absence of Yukawa couplings and partly for the possibility that
the scalar partners are naturally heavy.

In this framework, we consider the particular NMSSM case in which the mixing of the singlet
with doublet Higgs fields is absent in the CP-even sector and at tree level the lightest Higgs boson is exactly SM-like.
We analysed the predictions of this scenario
for a $M_h=125$ GeV Higgs  at the LHC. We show that the predictions
for the one-loop Higgs boson branching fractions and production rates in
$\gamma \gamma$ and $Z\gamma $ can
be sizably modified with respect to the FP SM model, allowing a
better fit  to present collider data. However, the tree-level Higgs decay
channels into $WW^*$ and $ZZ^*$ remain unaffected if the mixing between the
singlet and doublet Higgs fields is absent. Relaxing this last
condition, and so adding a new free parameter,
the Higgs coupling to weak gauge boson
$WW$ and $ZZ$ can be modified, and a suppression of the rates for
$h \to WW^* $ and $h \to ZZ^* $ with respect to the pure FP model expectations
can be achieved.

Finally, we would like to
stress that the FP NMSSM offers a new arena for SUSY phenomenology at the LHC.
In particular, most of the previous analyses on
SUSY particle searches should be revised in the light of the fact that
the large top-Yukawa coupling is absent or strongly suppressed.
In addition, the stringent constraints from Higgs mediated and other FCNC processes can  be
relaxed due to the absence or suppressed Yukawa couplings, allowing for a
light charged Higgs boson phenomenology at the LHC. Moreover,
the interplay between chiral and supersymmetry breaking
suggests that if there is a new strong dynamics at the TeV scale, as for instance
indicated by the large top-quark mass, this could also play
a role in the supersymmetry breaking mechanism, opening the way to a new and exciting
phenomenology at the LHC.

\section*{Appendix}
Here we provide the analytical expressions for the one-loop radiative
decays widths of
$h\to \gamma \gamma$ and $h\to Z \gamma$, where $h$ is
the lightest CP even Higgs boson, in the framework of pure
FP NMSSM model. Following the results of Refs.\cite{Djouadi:2005gj},
\cite{Djouadi:2005gi} we get
\bea
\Gamma(h\to \gamma \gamma) &=&
\frac{ \alpha^2 G_F M_h^3}{128 \sqrt{2}\pi^3}\Big|
g_{hWW}A^{\gamma}_1(\tau_W)+
\frac{M_W^2\lambda_{hH^{+}H^{-}}}{2c^2_W M_{H^{+}}^2}A^{\gamma}_0(\tau_{H^{+}})
\nonumber\\
&+&\sum_{i=1,2}\frac{2M_W}{M_{\chi^{+}_i}}g_{h\chi^{+}_i\chi^{-}_i}
A^{\gamma}_{1/2}(\tau_{\chi_i^{+}})\Big|^2\, ,
\label{Gammagg}
\eea

\bea
\Gamma(h\to Z \gamma) &=& \frac{\alpha G_F^2 M_W^2 M_h^3}{64\pi^4}
\left(1-\frac{M_Z^2}{M_h^2}\right)^3\Big|g_{hWW}
A^Z_1(\tau^{-1}_W,\lambda_W)
\nonumber\\
&+&\frac{M_W^2 v_{H^{\pm}}\lambda_{hH^{+}H^{-}}}{2c_W M_{H^{+}}^2}
A^Z_0(\tau^{-1}_{H^{+}},\lambda_{H^{+}})
\nonumber\\
&+&
\sum_{i=1,2;\, m=L,R}\frac{2M_W}{M_{\chi^{+}_i}}
g_{h\chi^{+}_i\chi_i^{-}}g^m_{Z\chi^{+}_i\chi^{-}_i}
A^Z_{1/2}(\tau^{-1}_{\chi_i^{+}},\lambda_{\chi_i^{+}})\, ,
\Big|^2\, ,
\label{GammaZg}
\eea
where $G_F$ is the Fermi constant, $\alpha$ the electromagnetic fine structure
constant, and the normalized Higgs and Z couplings appearing above
are given by \cite{Djouadi:2005gj}
\bea
g_{hWW}&=&\sin(\beta-\alpha)\, ,
\nonumber\\
g_{h\chi^{+}_i\chi^{-}_i}&=&\frac{1}{\sqrt{2} s_W}\Big(
-\sin\alpha\, V_{i1}U_{i2}+\cos\alpha\, V_{i2}U_{i1}\Big)\, ,
\nonumber\\
g^L_{Z\chi^{+}_i\chi^{-}_j}&=&\frac{1}{c_W}\left(s^2_W-\frac{1}{2}V_{i2}^2
-V_{i1}^2\right)\, ,
\nonumber\\
g^R_{Z\chi^{+}_i\chi^{-}_j}&=&\frac{1}{c_W}\left(s^2_W-\frac{1}{2}U_{i2}^2
-U_{i1}^2\right)\, ,
\nonumber\\
v_H^{\pm}&=&\frac{2c_W^2-1}{c_W}\, ,
\eea
where $c_W=\cos{\theta_W}$ and $s_W=\sin{\theta_W}$, with $\theta_W$
the Weinberg angle, $\lambda_{hH^{+}H^{-}}$ is given in eq. (\ref{eq:hHpHm}), $U_{ij}$ and $V_{ij}$ the matrix elements
of the corresponding $U,V$ matrices diagonalizing the chargino mass matrix
${\bf X}$ in Eq.(\ref{eq:charginomassmatrix}) as  $U {\bf X} V^{-1}$, and
$M_{\chi_i}$ the corresponding eigenvalues.

The other symbols appearing in the expressions of the widths in
Eqs.(\ref{Gammagg}) and (\ref{GammaZg}) are defined as
$\tau_i=M_h^2/(4M_i^2)$, $\lambda_i=4M_i^2/M_Z^2$, with
$i=W,H^{+},\chi^+_i$, while
the functions $A^\gamma_{(1/2,0,1)}(x)$ , and $A^Z_{(1/2,0,1)}(x,y)$
are given by \cite{Djouadi:2005gj},\cite{Djouadi:2005gi}
\begin{itemize}
\item for $h\to \gamma \gamma$
\bea
A^{\gamma}_{1/2}(x)&=& 2\left[x+\left(x-1\right) F(x)\,
\right] x^{-2}\, ,
\nonumber\\
\nonumber\\
A^{\gamma}_{0}(x)&=& -\left[x-F(x)\, \right]x^{-2}\, ,
\nonumber\\
\nonumber\\
A^{\gamma}_1(x)&=& -\left[2x^2+3x+3\left(2 x-1\right)F(x)\,
\right] x^{-2}\, ,
\eea
\end{itemize}
\begin{itemize}
\item for $h\to Z \gamma$
\bea
A^{Z}_{1/2}(x,y)&=& I_1(x,y)-I_2(x,y) \, ,
\nonumber\\
\nonumber\\
A^{Z}_{0}(x,y)&=& I_1(x,y)\, ,
\nonumber\\
\nonumber\\
A^Z_1(x,y)&=&c_W\Big(4\left(3-\frac{s^2_W}{c_W^2}\right) I_2(x,y)
        + \left[\left(1+\frac{2}{x}\right)\frac{s_W^2}{c_W^2}
        -\left(5+\frac{2}{x}\right)\right]I_1(x,y)\Big)\, ,
\label{loop}
\eea
\end{itemize}
where the functions $I_{1,2}(x,y)$ are given by
\bea
I_1(x,y)&=&\frac{xy}{2(x-y)}
+\frac{x^2 y^2}{2(x-y)^2}\left(F(x^{-1})-F(y^{-1})\right)+
    \frac{x^2y}{(x-y)^2}\left(G(x^{-1})-G(y^{-1})\right)\, ,
\nonumber\\
I_2(x,y)&=&-\frac{xy}{2(x-y)}\left(F(x^{-1})-F(y^{-1})\right)\, ,
\eea
with $F(x)=\left(\arcsin{\sqrt{x}}\right)^2$ for $x\le 1$, and
$G(x)=\sqrt{\frac{1-x}{x}}\arcsin{\sqrt{x}}$ for $x\le 1$.
The electromagnetic coupling constant $\alpha$, appearing in eqs.
(\ref{Gammagg}) and (\ref{GammaZg}), is evaluated at the scale
$q^2=0$,  since the final state photons
in the Higgs decays $H\to \gamma \gamma$ and $H\to Z \gamma$ are on shell.

\vskip 0.5in
\vbox{
\noindent{ {\bf Acknowledgements} } \\
\noindent
We thank A. Strumia for discussions. E.G. would like to thank the PH-TH division
of CERN for its kind hospitality during the preparation of this work.
This work was supported by the ESF grants  8090, 8499, 8943, MTT8, MTT59, MTT60, MJD140, JD164,  by the recurrent financing SF0690030s09 project
and by  the European Union through the European Regional Development Fund.
}

\end{document}